\documentclass[aps,prr,a4paper,superscriptaddress,twocolumn,showpacs,amsmath,amssymb,longbibliography]{revtex4-1}

\usepackage{graphicx,epsfig}
\usepackage[usenames]{color}

\usepackage[english]{babel}

\usepackage{dsfont}
\usepackage{bbold}

\usepackage{aliascnt}

\usepackage{csquotes}
\MakeOuterQuote{"}

\allowdisplaybreaks

\begin{document}

\newcommand{\E}{\mathcal{E}}
\newcommand{\Z}{\mathcal{Z}}
\newcommand{\B}{\mathcal{B}}
\newcommand{\Lag}{\mathcal{L}}
\newcommand{\M}{\mathcal{M}}
\newcommand{\N}{\mathcal{N}}
\newcommand{\U}{\mathcal{U}}
\newcommand{\R}{\mathcal{R}}
\newcommand{\F}{\mathcal{F}}
\newcommand{\V}{\mathcal{V}}
\newcommand{\C}{\mathcal{C}}
\newcommand{\I}{\mathcal{I}}
\newcommand{\s}{\sigma}
\newcommand{\up}{\uparrow}
\newcommand{\dw}{\downarrow}
\newcommand{\h}{\hat{\mathcal{H}}}
\newcommand{\himp}{\hat{h}}
\newcommand{\g}{\mathcal{G}^{-1}_0}
\newcommand{\D}{\mathcal{D}}
\newcommand{\A}{\mathcal{A}}

\newcommand{\projRi}{\hat{\mathcal{P}}_{i}^{\phantom{\dagger}}}
\newcommand{\projRidagger}{\hat{\mathcal{P}}_{i}^{{\dagger}}}
\newcommand{\proj}{\hat{\mathcal{P}}_G}

\newcommand{\K}{\textbf{k}}
\newcommand{\Q}{\textbf{q}}
\newcommand{\T}{\tau_{\ast}}
\newcommand{\io}{i\omega_n}
\newcommand{\eps}{\varepsilon}
\newcommand{\+}{\dag}
\newcommand{\su}{\uparrow}
\newcommand{\giu}{\downarrow}
\newcommand{\0}[1]{\textbf{#1}}
\newcommand{\ca}{c^{\phantom{\dagger}}}
\newcommand{\cc}{c^\dagger}
\newcommand{\fa}{f^{\phantom{\dagger}}}
\newcommand{\fc}{f^\dagger}
\newcommand{\aaa}{a^{\phantom{\dagger}}}
\newcommand{\aac}{a^\dagger}
\newcommand{\bba}{b^{\phantom{\dagger}}}
\newcommand{\bbc}{b^\dagger}
\newcommand{\da}{{d}^{\phantom{\dagger}}}
\newcommand{\dc}{{d}^\dagger}
\newcommand{\ha}{h^{\phantom{\dagger}}}
\newcommand{\hc}{h^\dagger}
\newcommand{\be}{\begin{equation}}
\newcommand{\ee}{\end{equation}}
\newcommand{\bea}{\begin{eqnarray}}
\newcommand{\eea}{\end{eqnarray}}
\newcommand{\ba}{\begin{eqnarray*}}
\newcommand{\ea}{\end{eqnarray*}}
\newcommand{\dagga}{{\phantom{\dagger}}}
\newcommand{\bR}{\mathbf{R}}
\newcommand{\bQ}{\mathbf{Q}}
\newcommand{\bq}{\mathbf{q}}
\newcommand{\bqp}{\mathbf{q'}}
\newcommand{\bk}{\mathbf{k}}
\newcommand{\bh}{\mathbf{h}}
\newcommand{\bkp}{\mathbf{k'}}
\newcommand{\bp}{\mathbf{p}}
\newcommand{\bL}{\mathbf{L}}
\newcommand{\bRp}{\mathbf{R'}}
\newcommand{\bx}{\mathbf{x}}
\newcommand{\by}{\mathbf{y}}
\newcommand{\bz}{\mathbf{z}}
\newcommand{\br}{\mathbf{r}}
\newcommand{\Ima}{{\Im m}}
\newcommand{\Rea}{{\Re e}}
\newcommand{\Pj}[2]{|#1\rangle\langle #2|}
\newcommand{\ket}[1]{\vert#1\rangle}
\newcommand{\bra}[1]{\langle#1\vert}
\newcommand{\setof}[1]{\left\{#1\right\}}
\newcommand{\fract}[2]{\frac{\displaystyle #1}{\displaystyle #2}}
\newcommand{\Av}[2]{\langle #1|\,#2\,|#1\rangle}
\newcommand{\av}[1]{\langle #1 \rangle}
\newcommand{\Mel}[3]{\langle #1|#2\,|#3\rangle}
\newcommand{\Avs}[1]{\langle \,#1\,\rangle_0}
\newcommand{\eqn}[1]{(\ref{#1})}
\newcommand{\Tr}{\mathrm{Tr}}

\newcommand{\Vb}{\bar{\mathcal{V}}}
\newcommand{\Vd}{\Delta\mathcal{V}}
\def\P{\mathcal{P}}
\newcommand{\Pb}{\bar{P}_{02}}
\newcommand{\Pd}{\Delta P_{02}}
\def\t{\theta_{02}}
\newcommand{\tb}{\bar{\theta}_{02}}
\newcommand{\td}{\Delta \theta_{02}}
\newcommand{\Rb}{\bar{R}}
\newcommand{\Rd}{\Delta R}
\newcommand{\ocrev}[1]{{\color{cyan}{#1}}}
\newcommand{\occom}[1]{{\color{red}{#1}}}


\title{Derivation of the Ghost Gutzwiller Approximation from Quantum Embedding principles: the Ghost Density Matrix Embedding Theory}

\author{Nicola Lanat\`a}
\altaffiliation{Corresponding author: nxlsps@rit.edu}
\affiliation{School of Physics and Astronomy, Rochester Institute of Technology,
84 Lomb Memorial Drive, Rochester, New York 14623, USA}
\affiliation{Center for Computational Quantum Physics, Flatiron Institute, New York, New York 10010, USA}

\date{\today}

\begin{abstract}
Establishing the underlying links between the diverse landscape of theoretical frameworks for simulating strongly correlated matter is crucial for advancing our understanding of these systems.
In this work, we focus on the Ghost Gutzwiller Approximation (gGA), an extension of the Gutzwiller Approximation (GA) based on the variational principle.
We derive a framework called  "Ghost Density Matrix Embedding Theory" (gDMET) from quantum embedding (QE) principles similar to those in Density Matrix Embedding Theory (DMET), which reproduces the gGA equations for multi-orbital Hubbard models with a simpler implementation. 
This derivation highlights the crucial role of the ghost degrees of freedom, not only as an extension to the GA, but also as the key element in establishing a consistent conceptual connection between DMET and the gGA.
This connection further elucidates how gGA overcomes the systematic accuracy limitations of standard GA and achieves results comparable to Dynamical Mean Field Theory (DMFT).
Furthermore, it offers an alternative interpretation of the gGA equations, fostering new ideas and generalizations.
\end{abstract}

\maketitle

\section{Introduction}

Theoretical frameworks based on QE principles~\cite{Kotliar-Science, quantum-embedding-review} have emerged as powerful tools for studying strongly correlated matter. Among these, DMFT~\cite{DMFT,dmft_book,Held-review-DMFT,Anisimov_DMFT,LDA+U+DMFT} is a well-known and widely used method. 
Other methods, such as the GA~\cite{Gutzwiller3,GA-infinite-dim,Fang,Ho,Gmethod,Our-PRX,Bunemann,Attaccalite}, its recent extension gGA~\cite{Ghost-GA,ALM_g-GA,ghost-Guerci}, the Rotationally Invariant Slave Boson (RISB) theory~\cite{Georges-RISB,rotationally-invariant_SB,Lanata2016}, the Slave Spin theory~\cite{Slave-spin-deMedici,Slave-spin-Qimiao}, and DMET~\cite{DMET,DMET-qchem,Bulik-DMET,dmet-new-2,dmet-new-3,dmet-new-4,dmet-new-5,dmet-new-6,dmet-thesis}, have also made significant contributions. In this work, we focus on GA and gGA, both of which are based on the variational principle and, as DMFT, on the limit of infinite dimensionality~\cite{Gutzwiller3,GA-infinite-dim}.

The key idea underlying gGA is to expand the GA variational space by incorporating auxiliary "ghost" fermionic degrees of freedom ---which is a common theme with different frameworks such as: extensions to DMET~\cite{Booth-ghost}, matrix product states and projected entangled pair states~\cite{Vaestrate-VRG}, the ancilla qubit techique~\cite{ghost-Subir}, and recent extensions of neural network states~\cite{ghost-NeuralNetworks}. It also presents suggestive analogies with the physical concepts of "hidden Fermion"~\cite{ghost-HF} and "hidden Fermi liquid"~\cite{Ghost-Anderson}.

The gGA variational extension allows achieving an accuracy comparable to that of DMFT, but with a substantially lower computational cost~\cite{Ghost-GA,ALM_g-GA,comp_gGA-DMFT,ghost-Carlos}. Additionally, the gGA, like the GA, can be reformulated using a RISB perspective~\cite{Kotliar-Ruckenstein,equivalence_GA-SB,lanata-barone-fabrizio,Operatorial-gRISB}, providing us with an exact reformulation of the many-body problem that reduces to gGA at the mean field level. This alternative formulation may pave the way to develop practical implementations for adding systematically quantum-fluctuation corrections towards the exact solution.

In Refs.~\cite{Our-PRX,Ghost-GA,ALM_g-GA} it was shown that the GA and the gGA can be both formulated using a typical QE algorithmic structure, analogous to DMET. This structure involves the recursive computation of an Embedding Hamiltonian's (EH) ground state for each correlated fragment of the system. In practice, this approach offers new opportunities to reduce the computational costs of GA and gGA ---e.g., by employing density matrix renormalization group~\cite{DMRG-original-White-PRL}, variational quantum eigensolvers~\cite{VQE-review,YY-qc,Rogers2023ErrorMitigation}, or other classical methods~\cite{Coupled-cluster-RevModPhys,GRG-Lanata}, to compute the EH's ground state.
Subsequently, the comparison between the GA and DMET
equations was also discussed in Refs.~\cite{dmet-risb-1,dmet-risb-2}, where it was noted that the main mathematical difference between the DMET and the GA, as formulated in Refs.~\cite{Our-PRX,Lanata2016} using a QE structure, is that the latter involves variational parameters encoding the quasi-particle spectral weights of the correlated degrees of freedom, which are effectively set to 1 in DMET.

The mathematical similarities between GA, gGA, and DMET algorithms, outlined above, suggest a possible underlying physical connection between these methods. However, such a connection has not been established yet.

In this paper we address this issue by deriving the gDMET: a QE method based on principles similar to those of DMET, possessing self-consistency conditions mathematically equivalent to those found in the gGA variational-energy minimization framework~\cite{Ghost-GA,ALM_g-GA}.
This clear correspondence between gGA and gDMET offers a valuable alternative perspective on interpreting the physical implications of the resulting equations, 
and introduces a practical advantage with a simpler implementation. 
Furthermore, as our approach yields non-arbitrary self-consistency conditions consistent with the variational principle in the infinite-dimensional limit, it could serve as a guide for developing new QE methods with enhanced accuracy and broader applicability, e.g.,  for systems with non-local interactions~\cite{Nonlocal-dmft-1,Nonlocal-dmft-2,Nonlocal-dmft-3,Nonlocal-dmft-4,Nonlocal-boh,Nonlocal-Lanata}, at finite temperature~\cite{finiteT-GA-Wang,finiteT-GA-fabrizio,finiteT-GA-lanata,finiteT-DMET}, and out of equilibrium~\cite{tdGA-Seibold,tdGA-Michele,tdGA-Lanata,td-gGA-Lanata,td-DMET,tdDMET-Garnet}.

\section{Derivation of $\text{g}$DMET: the $\text{g}$GA from QE self-consistency principles}

Let us consider a generic multi-orbital Fermi-Hubbard Hamiltonian represented as follows:
\begin{align}
    \hat{H}&=\sum_{i=1}^{\mathcal{N}}\hat{H}_{loc}^i[\cc_{i\alpha},\ca_{i\alpha}]+\sum_{i\neq j}\hat{T}_{ij}
    \label{H}
    \\
    \hat{T}_{ij}&=\sum_{\alpha=1}^{\nu_i}\sum_{\beta=1}^{\nu_j}[t_{ij}]_{\alpha\beta}\,\cc_{i\alpha}\ca_{j\beta}
    \,,
\end{align}
where $i$ and $j$ label the fragments of the system, $\hat{H}_{loc}^i$ is a generic operator lying within the $i$ fragment (i.e., constructed with $\cc_{i\alpha},\ca_{i\alpha}$), including both one-body and two-body contributions, and $\alpha$ labels all Fermionic modes within each fragment.

The key idea underlying QE frameworks is to describe the interaction of each fragment with its environment in terms of an EH, consisting of the fragment and an entangled quantum bath. 
In principle, it can be demonstrated that the fragments can always be exactly embedded by baths no larger than the fragments themselves. However, this result is purely formal. 
In DMET a practical approximation to the bath of the EH is built from a one-body state, which is generally constructed as the ground state of an auxiliary one-body Hamiltonian $\hat{H}_*$, determined by appropriate self-consistency conditions.

The goal of this section is to develop the gGA from QE self-consistency principles reminiscent of DMET. 
Like gGA, our construction will involve an effective one-body Hamiltonian $\hat{H}_*$ featuring auxiliary Fermionic degrees of freedom, that will serve to enrich the description of many-body effects compared to classical DMET frameworks. 

\begin{figure}
    \centering
    \includegraphics[width=0.45\textwidth]{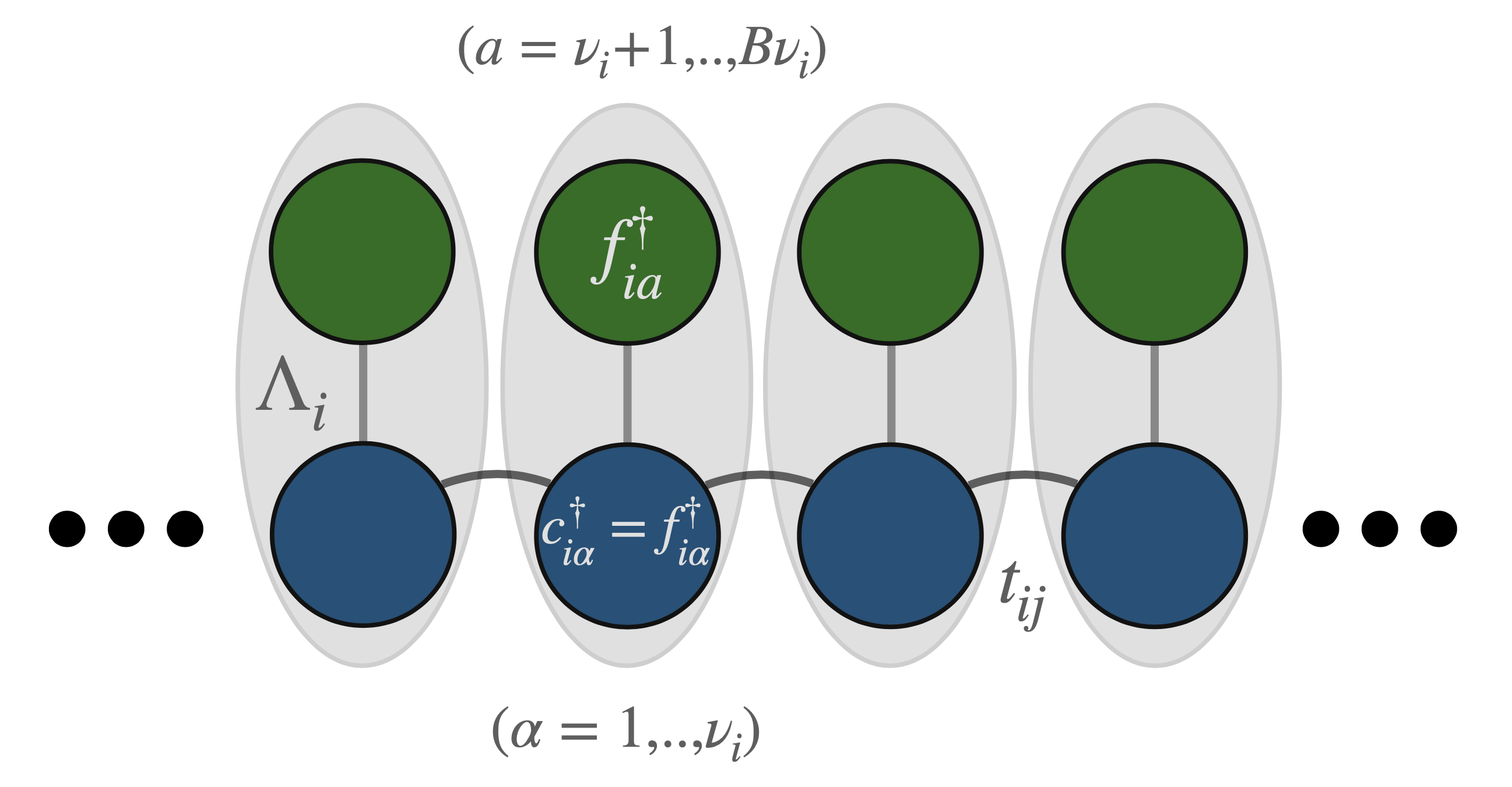}
    \caption{Schematic 1-dimensional representation of the effective Hamiltonian $\hat{H}_*$ in the ideal scenario described in Sec.~\ref{Hqp-sec}. The physical degrees of freedom $\cc_{i\alpha}=\fc_{i\alpha}$, $\alpha=1,..,\nu_i$, are represented by blue circles. The additional ghost modes $\fc_{ia}$, $a=\nu_i\!+\!1,..,B\nu_i$, are represented by green circles.}
   \label{Figure1}
\end{figure}

\subsection{The gDMET quasiparticle Hamiltonian}\label{Hqp-sec}

In our approach, we construct the effective Hamiltonian $\hat{H}_*$, also called "quasiparticle Hamiltonian," as in the gGA literature:
\begin{align}
    \hat{H}_*[\R,\Lambda]&=\sum_{i=1}^{\mathcal{N}}\sum_{a,b=1}^{{B}\nu_i}[\Lambda_i]_{ab}\,\fc_{ia}\fa_{ib}
    +\sum_{i\neq j}\hat{T}_{ij}
    \nonumber\\
    \hat{T}_{ij}&=\sum_{a=1}^{B\nu_i}\sum_{b=1}^{B\nu_j}[\R_it_{ij}\R^\dagger_j]_{ab}\,\fc_{ia}\fa_{jb}
    \,,
    \label{Hqp}
\end{align}
where $\R=(\R_1,..,\R_\mathcal{N})$, $\Lambda=(\Lambda_1,..,\Lambda_\mathcal{N})$, we assume that $B>1$ 
is an odd number, and we introduce modes $\fc_{ia}$, with original Fermionic modes $\cc_{i\alpha}$ expressed as linear combinations:
\begin{align}
    \cc_{i\alpha}=\sum_{a=1}^{{B}\nu_i}[\R_i]_{a\alpha}\fc_{ia}
    \;\quad (\alpha=1,..,\nu_i)
    \,.
    \label{cfR}
\end{align}
Similarly to all DMET implementations, the entries of the parameters $\Lambda_i$ and $\R_i$ characterizing
$\hat{H}_*$ are initially unspecified, and will be determined self-consistently.

For Eq.~\eqref{cfR} to be consistent and provide $\cc_{i\alpha}$ and $\fc_{ia}$ modes both satisfying canonical anticommutation rules:
\begin{align}
\{\fa_{ia},\fc_{jb}\}&=\delta_{ij}\delta_{ab}
\\
\{\ca_{i\alpha},\cc_{j\beta}\}&=\delta_{ij}\delta_{\alpha\beta}\,,     
\end{align}
the condition $\R_i^\dagger \R_i=\mathbf{1}$ (where $\mathbf{1}$ is the identity matrix) has to hold true.
When this condition is satisfied exactly, the modes $\fc_{ia}$ can be chosen in such a way that $[\R_i]_{a\alpha}=\delta_{a\alpha}$ $\forall\,a,\alpha\leq \nu_i$ and $[\R_i]_{a\alpha}=0$ otherwise.
This ideal scenario,
corresponding to the system represented in Fig.~1 is useful for interpreting the gGA equations from a DMET perspective.
Specifically, it allows interpreting $\hat{H}_*$ as an effective Hamiltonian approximating many-body interactions between fragments and their environments using one-body operators; with "hopping" (non-local) terms retained as in the original Hamiltonian $\hat{H}$, and new "ghost" or "ancilla" Fermionic degrees of freedom introduced locally in all fragments to enrich the approximate description of many-body effects induced by local interacting terms of $\hat{H}$.

For later convenience, we rewrite Eq.~\eqref{Hqp} as follows:
\begin{align}
    \hat{H}_*[\R,\Lambda]
    &=\sum_{i,j=1}^{\mathcal{N}}
    [\Pi_i h_* \Pi_j]_{ab}\,\fc_{ia}\fa_{jb}
    \,,
    \label{Hqp2}
\end{align}
where we introduced the matrix:
\begin{align}
    h_* = \begin{pmatrix}
    \Lambda_1  &\R_1t_{12}\R^\dagger_2& \dots & \R_{1}t_{1 \mathcal{N}}\R^\dagger_{\mathcal{N}} \\
    \R_2t_{21}\R^\dagger_1 & \Lambda_2 & \dots & \vdots \\
    \vdots  & \vdots  & \ddots & \vdots \\
    \R_{\mathcal{N}1}t_{\mathcal{N} 1}\R^\dagger_{1} & \dots& \dots & \Lambda_{\mathcal{N}}
  \end{pmatrix}
\end{align}
and the projectors over the degrees of freedom corresponding to each fragment:
\begin{align}
   \label{Proj}
   \Pi_i = \begin{pmatrix}
     \delta_{i1}\left[\mathbf{1}\right]_{B{\nu}_1\times B{\nu}_1} & \dots & \mathbf{0} \\
     \vdots  & \ddots & \vdots \\
    \mathbf{0}  &  \dots & \delta_{iM}\left[\mathbf{1}\right]_{B{\nu}_M\times B{\nu}_M}
  \end{pmatrix}
  \,,
\end{align}
where $\left[\mathbf{1}\right]_{n\times n}$ is the $n\times n$ identity matrix.

\begin{figure}
    \centering
    \includegraphics[width=0.35\textwidth]{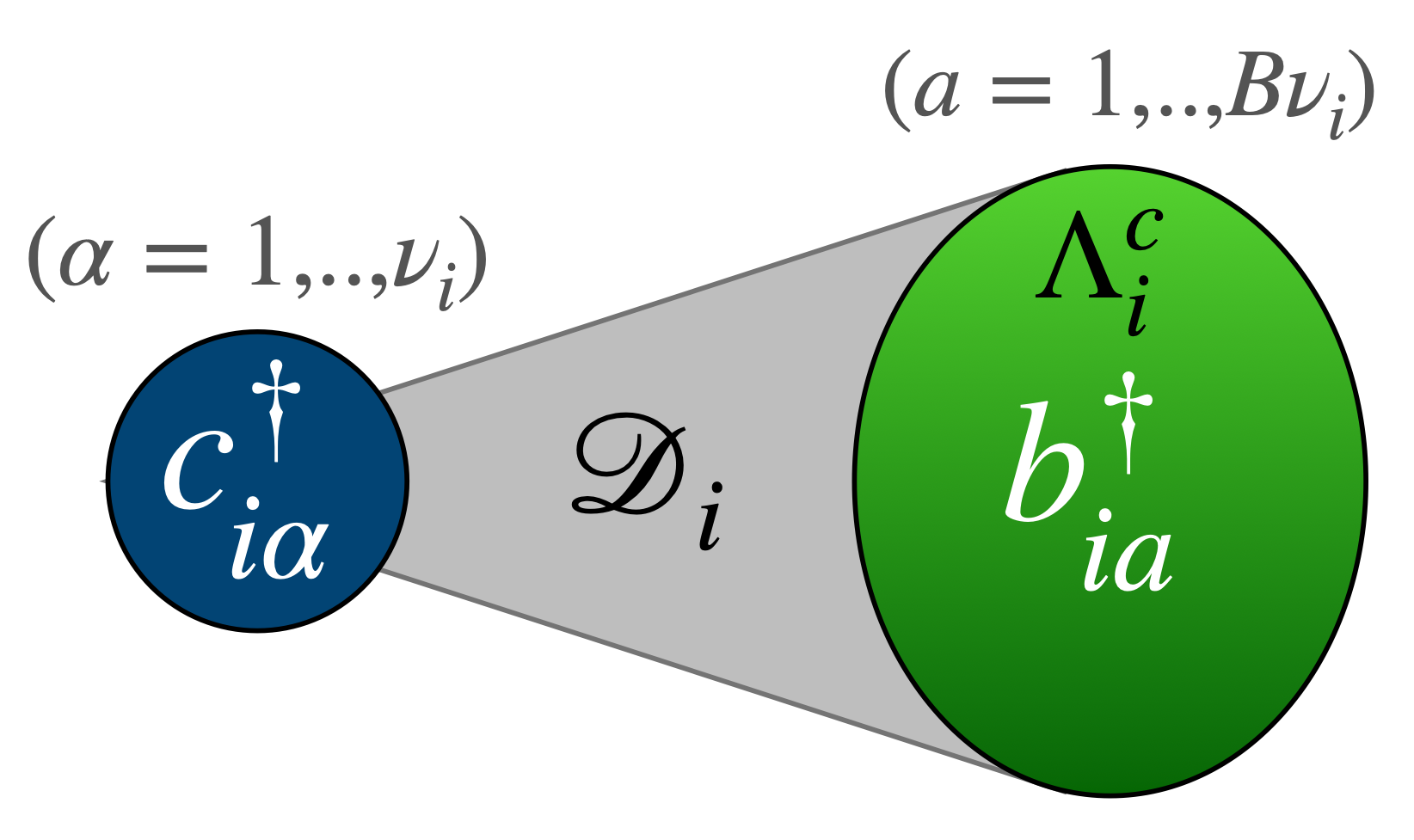}
    \caption{Schematic representation of the EH $\hat{H}^i_{emb}$. The fragment degrees of freedom $\cc_{i\alpha}$, $\alpha=1,..,\nu_i$, are represented by a blue circle, while the additional bath modes $\fc_{ia}$, $a=1,..,B\nu_i$, are represented by a green ellipse. The matrix $\Lambda^c_i$ encodes the bath degrees of freedom and the ghost modes, while the matrix $\D_i$ encodes the hybridization with the fragment.}
   \label{Figure2}
\end{figure}

\subsection{Construction of the Embedding Hamiltonians}

In this section, we outline the construction of embedding Hamiltonians for individual fragments of $\hat{H}$, leveraging a general theorem based on the Schmidt decomposition, here provided in the supplemental material for completeness~\cite{supplemental_material}.
This theorem, a standard result in DMET frameworks~\cite{dmet-guide}, allows expressing the active part of the ground state $\ket{\Psi_0}$ of $\hat{H}_*$ (see Eq.~\eqref{Hqp}) for each fragment $i$.
%
Specifically, the theorem demonstrates that the active part corresponds to the ground state of a one-body EH involving $B\nu_i$ fragment degrees of freedom, denoted as $\fc{ia}$, and $B\nu_i$ "bath" degrees of freedom entangled with it, denoted as $\bbc_{ia}$, given by the following equation:
%
\begin{align}
    b^\dagger_{ia}&=\sum_{j=1}^{\mathcal{N}}\sum_{b=1}^{B\nu_j} [\B^j_{i}]_{b a}\, \fc_{jb}\;\quad (a=1,..,B\nu_i)
    \label{Bn-recap}
    \,,
\end{align}
where the entries of the matrices $\B^j_i$ are given by the following equation:
\begin{align}
    [\B_i^j]_{ba} 
    =(1-\delta_{ij})
    \left[\Pi_jf(h_*)\Pi_i\frac{1}{\sqrt{^t\Delta_i(\mathbf{1}- ^t\!\Delta_i)}}\right]_{ba}
    \,,
\end{align}
where:
\begin{align}
    [\Delta_i]_{ab}=\Av{\Psi_0}{\fc_{ia} \fa_{ib}}\;\quad(a,b=1,..,B\nu_i)
\end{align}
and $^t\Delta_i$ indicates the 
transpose of $\Delta_i$.
For later convenience, we define also the following matrix:
\begin{align}
    \B_i &= \sum_{j=1}^{\mathcal{N}} \B_i^j
    \,.
\end{align}

As explained in the supplemental material~\cite{supplemental_material}, the resulting effective one-body embedding Hamiltonian associated to each fragment $i$ is the following:
\begin{align}
    \hat{H}^i_0&=
    \sum_{a,b=1}^{{B}\nu_i}[\Lambda_i]_{ab}\,\fc_{ia}\fa_{ib}
    -\sum_{a,b=1}^{B\nu_i}[\Lambda^c_i]_{ab}
    \,\bbc_{ia}\bba_{ib}
    \nonumber
    \\
    &+\sum_{a,b=1}^{B\nu_i}
    \left(
    \left[\D_i^0 \right]_{ba}
    \,\fc_{ia}\bba_{ib}
    +\text{H.c.}
    \right)
    \nonumber\\
    &=\sum_{a,b=1}^{{B}\nu_i}[\Lambda_i]_{ab}\,\fc_{ia}\fa_{ib}
    -\sum_{a,b=1}^{B\nu_i}[\Lambda^c_i]_{ab}
    \,\bbc_{ia}\bba_{ib}
    \nonumber
    \\
    &+\sum_{a=1}^{B{\nu}_i}\sum_{\alpha=1}^{\nu_i} 
    \left(
    \left[\D_i \right]_{b\alpha}
    \,\cc_{i\alpha}\bba_{ib}
    +\text{H.c.}
    \right)
    \label{Hi0}
    \,,
\end{align}
where:
\begin{align}
    [\Lambda^c_i]_{ab}&=
    -[\Pi_i \B_i^\dagger h_* \B_i\Pi_i]_{ab}
    \\
    \left[\D_i^0 \right]_{ba}&=
    \left[\Pi_ih_*\B_i\right]_{ab}=\sum_j
    \left[\R_it_{ij}\R^\dagger_j\B^j_i\right]_{a b}
    \\
    \left[\D_i \right]_{ba}&=\sum_j
    \left[t_{ij}\R^\dagger_j\B^j_i\right]_{a b}
    \,,
\end{align}
and in the last step of Eq.~\eqref{Hi0} we used Eq.~\eqref{cfR}.

Given $\hat{H}^i_0$, we construct the following approximation to the actual EH of each impurity for the original interacting Hamiltonian $\hat{H}$ (Eq.~\eqref{H}) as follows:
\begin{align}
    \hat{H}^i_{emb}&=\hat{H}^i_0-\sum_{a,b=1}^{{B}\nu_i}[\Lambda_i]_{ab}\,\fc_{ia}\fa_{ib}
    +\hat{H}_{loc}^i[\cc_{i\alpha},\ca_{i\alpha}]
    \nonumber\\
    &=\hat{H}_{loc}^i[\cc_{i\alpha},\ca_{i\alpha}]
    -\sum_{a,b=1}^{B\nu_i}[\Lambda^c_i]_{ab}
    \,\bbc_{ia}\bba_{ib}
    \nonumber
    \\
    &+\sum_{a=1}^{B{\nu}_i}\sum_{\alpha=1}^{\nu_i}
    \left(
    \left[\D_i \right]_{b\alpha}
    \,\cc_{i\alpha}\bba_{ib}
    +\text{H.c.}
    \right)
    \,,
    \label{EH}
\end{align}
which is represented schematically in Fig.~\ref{Figure2}.

This working hypothesis, which will enable us to recover the gGA equations, can be physically motivated by noting that the fragment portion of the EH is already known to be $\hat{H}_{loc}^i$. Therefore, that part does not need to be approximated.

Note that $\hat{H}^i_{emb}$ involves explicitly only the $\nu_i$ fragment modes $\cc_{i\alpha}$ and $B\nu_i$ bath modes $\bbc_{ia}$. However, it is crucial to understand that, by construction, $\hat{H}^i_{emb}$ exists within the extended Hilbert space containing all $B\nu_i$ impurity degrees of freedom $f^\dagger_{ia}$, not only the physical ones. As explained in the supplemental material~\cite{supplemental_material}, the EH construction requires that the total EH system (including the additional $(B-1)\nu_i$ auxiliary modes) is half-filled, containing $B\nu_i$ Fermions in total.
This requirement introduces the challenge of assigning the available $B\nu_i$ Fermions between the $(B-1)\nu_i$ auxiliary modes and the physically relevant portion of the EH, corresponding to the $(B+1)\nu_i$ modes explicitly appearing in Eq.~\eqref{EH}. The same question arises within the context of gGA~\cite{Ghost-GA,ALM_g-GA,supplemental_material}, where different choices for assigning Fermions may lead to different variational states, among which the physical solution is the one resulting in the lowest variational energy.

In previous work it was found that the physically relevant solution is generally obtained when the physically relevant portion of the EH, corresponding to the $(B+1)\nu_i$ modes explicitly appearing in Eq.~\eqref{EH},
is itself half filled, i.e., it contains $\nu_i(B+1)/2$ Fermions (which is integer for odd $B$, as we previously assumed in Sec.~\ref{Hqp-sec}).
Therefore, from now on we are going to call $\ket{\Phi_i}$ the ground state of $\hat{H}_{loc}^i$ within the subspace with $\nu_i(B+1)/2$ Fermions.

\subsection{Self-consistency conditions}

We now postulate the following self-consistency conditions for determining the parameters $\R_i$ and $\Lambda_i$:
\begin{align}
    \Av{\Psi_0[\R,\Lambda]}{\bbc_{ia}\bba_{ib}}&=\Av{\Phi_i}{\bbc_{ia}\bba_{ib}}
    \label{SC-bath}
    \\
    \Av{\Psi_0[\R,\Lambda]}{\cc_{i\alpha}\bba_{ib}}&=\Av{\Phi_i}{\cc_{i\alpha}\bba_{ib}}
    \label{SC-hybr}
    \,.
\end{align}

This working hypothesis, that will allow us to recover the gGA equations, can be physically motivated by
observing that the role of $\ket{\Psi_0}$ is to approximate the environment of each fragment and its interaction with the fragment itself, which are directly related with Eqs.~\eqref{SC-bath} and
\eqref{SC-hybr}, respectively.

Note that, as shown in the supplemental material~\cite{supplemental_material}, the left hand side of Eq.~\eqref{SC-bath} is given by:
\begin{align}
    \Av{\Psi_0[\R,\Lambda]}{\bbc_{ia}\bba_{ib}}=
    [\mathbf{1}-\Delta_i]_{ab}
    \,.
\end{align}

Similarly, the left hand side of Eq.~\eqref{SC-hybr} is given by the following equation:
\begin{align}
    &\Av{\Psi_0[\R,\Lambda]}{\cc_{i\alpha}\bba_{ib}}
    \nonumber\\
    &\qquad=\sum_{a=1}^{B\nu_i} [\R_{i}]_{a\alpha}
    \Av{\Psi_0[\R,\Lambda]}{\fc_{ia}\bba_{ib}}
    \nonumber\\
    &\qquad=\sum_{a=1}^{B\nu_i} [\R_{i}]_{a\alpha}
    [\Delta_i(\mathbf{1}-\Delta_i)]^{\frac{1}{2}}_{ab}
    \,.
\end{align}

\subsection{Summary}
In summary, the solution to the gDMET quantum embedding problem is given by the following identities:
\begin{align}
    \hat{H}_*[\R,\Lambda]\ket{\Psi_0}&=E_0\ket{\Psi_0}
    \label{Hqp-summary}
    \\
    [\Delta_i]_{ab}&=\Av{\Psi_0}{\fc_{ia} \fa_{ib}}
    \label{Delta-summary}
    \\
    \B_i^j 
    &=(1\!-\!\delta_{ij}) \, 
    \Pi_jf(h_*)\Pi_i\frac{1}{\sqrt{^t\Delta_i(\mathbf{1}-^t\!\Delta_i)}}
    \\
    \B_i &= \sum_{j=1}^{\mathcal{N}} \B_i^j
    \\
    [\D_i]_{b\alpha}&=\sum_{j=1}^{\mathcal{N}}
    \left[t_{ij}\R^\dagger_j\B^j_i\right]_{\alpha b}
    \label{D-summary}
    \\
    [\Lambda^c_i]_{ab}&=-[\Pi_i \B_i^\dagger h_* \B_i\Pi_i]_{ab}
    \label{Lc-summary}
    \\
    \hat{H}^i_{emb}\ket{\Phi_i}&=E^c_i\ket{\Phi_i}
    \label{Hemb-summary}
    \\
    \Av{\Phi_i}{\bbc_{ia}\bba_{ib}}&=[\mathbf{1}-\Delta_i]_{ab}
    \label{SC-bath-summary}
    \\
    \Av{\Phi_i}{\cc_{i\alpha}\bba_{ib}}&=\sum_{a=1}^{B\nu_i} [\R_{i}]_{a\alpha}
    [\Delta_i(\mathbf{1}-\Delta_i)]^{\frac{1}{2}}_{ab}
    \label{SC-hybr-summary}
    \,,
\end{align}
where the embedding parameters $\Lambda^c_i$ and $\Delta_i$,
characterizing the EH $\hat{H}^i_{emb}$ of the fragments, depend on $\R_i$ and $\Lambda_i$ through the steps described above; Eq.~\eqref{Hemb-summary} consists
in computing the ground state of $\hat{H}^i_{emb}$;
and Eqs.~\eqref{SC-bath-summary},\eqref{SC-hybr-summary} are
the self-consistency conditions to be satisfied for determining
$\R_i$ and $\Lambda_i$.

In the supplemental material, we provide a proof that the gDMET equations presented above are indeed mathematically equivalent to the gGA equations~\cite{supplemental_material}. This equivalence allows for a complementary physical interpretation. 
Furthermore, our approach introduces a practical advantage, as Eq.~\eqref{Lc-summary} offers a simpler method for calculating $\Lambda^c_i$ compared to the expression used in previous GA/gGA implementations~\cite{supplemental_material}.

A key mathematical distinction of our approach, compared to other DMET implementations proposed in the literature, is that Eqs.~\eqref{Hqp-summary}-\eqref{SC-hybr-summary} do not form an under-determined system, but can be exactly satisfied simultaneously. The primary reason is that the combined number of independent entries of $\Lambda_i$ and $\R_i$ is equal to the number of independent EH density matrix elements appearing in the left-hand sides of Eqs.~\eqref{SC-bath-summary},\eqref{SC-hybr-summary}. In classic DMET frameworks, however, the matrices $\R_i$ are not considered as free parameters, and the parameters $\Lambda_i$ are insufficient for exactly satisfying the QE self-consistency conditions. The standard DMET approach to tackle this problem is to address the self-consistency conditions only approximately, with respect to an arbitrary notion of distance.
On the other hand, the QE framework derived in this work relies on the assumption that the matrices $\R_i$ satisfy the condition $\R^\dagger_i \R_i=\mathbf{1}$, which should therefore be interpreted as a "sanity check" of the theory, but is not generally exactly verified.
The discrepancy between the ideal assumption of $\R^\dagger_i \R_i=\mathbf{1}$ and the actual matrix conditions found in practice hints at a common underlying physical reason that is relevant to both our approach and the classic DMET frameworks.

From a physical standpoint, this common underlying physical reason can be traced back to the role of the ghost degrees of freedom. In classic GA (i.e., for $B=1$, corresponding to the limiting case without ghost Fermionic degrees of freedom), $\R^\dagger_i \R_i=\Z_i$ represents the quasi-particle weight of the $i$ degrees of freedom. Since generally the eigenvalues of $\Z_i$ are smaller than $1$, the QE procedure above becomes less justified in the correlated regime from a DMET perspective (while it remains perfectly justified from the GA variational perspective). In contrast, gGA incorporates ghost Fermionic degrees of freedom, where $\R^\dagger_i \R_i$ represents the entire spectral weight of the $i$ degrees of freedom, including both the quasiparticle weight and the contribution of the Hubbard bands. Therefore, in gGA one generally finds that $\R^\dagger_i \R_i\sim \mathbf{1}$ in all physical regimes, ranging from the weakly correlated to the strongly correlated. This fact can be interpreted as an a-posteriori indication that the additional ghost degrees of freedom allow us to capture with higher precision the many-body effects induced by local interacting terms of $\hat{H}$, justifying our procedure.

In light of the above discussion, we argue that the ghost degrees of freedom play a pivotal role in unifying the DMET and gGA perspectives, offering a common framework that successfully captures many-body effects, while adhering to the principles of both approaches.

\section{Gauge freedom and  QE gauge from Singular Value Decomposition}

In this section, we discuss the gauge invariance of the equations derived in our paper. By inspection, one can verify that Eqs.~\eqref{Delta-summary}-\eqref{SC-hybr-summary} are invariant with respect to the following set of gauge transformations:
\begin{align}
    \ket{\Psi_0} &\rightarrow \hat{\U}^{\dagger}\left(\theta_1,..,\theta_{\mathcal{N}}\right)\ket{\Psi_0} 
    \label{g1}
    \\
    \ket{\Phi_i} &\rightarrow \hat{U}_i^{\dagger}\left(\theta_i\right)\ket{\Phi_i} \label{gaugePhi}\\
    \R_i &\rightarrow u^{\dagger}_i\left(\theta_i\right)\R_i \\
    \D_i &\rightarrow {}^t u\left(\theta_i\right)\D_i \\
    \Delta_i &\rightarrow {}^t u_i\left(\theta_i\right)\Delta_i {}^t u_i^{\dagger}\left(\theta_i\right) \\
    \Lambda_i &\rightarrow u_i^{\dagger}\left(\theta_i\right)\Lambda_i u_i \left(\theta_i\right) \\
    \Lambda_i^c &\rightarrow u_i^{\dagger}\left(\theta_i\right) \Lambda_i^c u_i\left(\theta_i\right)
    \label{g7}
    \,,
\end{align}
where $\theta_i$ are Hermitian matrices of Lie parameters and:
\begin{align}
    u_i\left(\theta_i\right) &= e^{i\theta_i} \\
    \hat{U}_i\left(\theta_i\right) &= e^{i\sum_{a,b=1}^{B{\nu}_i}\left[\theta_i\right]_{ab} \bbc_{ia}\bba_{ib}} \\
    \hat{\U}\left(\theta_1,..,\theta_{\mathcal{N}}\right) &= e^{i\sum_{i}\sum_{a,b=1}^{B{\nu}_i}\left[\theta_i\right]_{ab} \fc_{ia}\fa_{ib}}
\,.
\end{align}

In the context of our derivation of the QE equations, the invariance under the given set of gauge transformations stems from the fact that the choice of the $\fa_{ia}$ modes is arbitrary, as long as they can span the physical $\ca_{i\alpha}$ modes. The term "gauge" refers to the idea that such a basis choice does not impact physical quantities, since all parameters related by this group of transformations are effectively equivalent.
In particular, note that Eq.~\eqref{gaugePhi} corresponds to applying any unitary transformation to the bath of the effective Hamiltonian, which does not affect the expectation values of fragment observables in the EH (as in DMFT).

Importantly, as discussed in Sec.~\ref{Hqp-sec}, if the identity $\R_i^\dagger \R_i=\mathbf{1}$ was exactly satisfied, we would be able to choose a gauge that corresponds to the ideal scenario depicted in Fig. 1 ---which was used to justify our QE procedure.
On the other hand, even when $\R_i^\dagger \R_i=\mathbf{1}$ is verified only approximately, we can leverage the gauge freedom to transform $\R_i$ into a form that closely resembles the ideal scenario depicted in Fig. 1 (which we are going to refer to as the "QE gauge"). To illustrate this point, we will present an argument based on the singular value decomposition (SVD) of $\R_i$.

The SVD theorem states that it is always possible to express a rectangular matrix such as $\R_i$ as follows:
\begin{equation}
    \R_i = U_i \Sigma_i V_i^\dagger\,,
\end{equation} 
where $U_i$ is a $B\nu_i\times B\nu_i$ matrix, $V_i$ is a $\nu_i\times \nu_i$ unitary matrix and 
$\Sigma_i$ is a $B\nu_i\times \nu_i$ diagonal matrix, where $[\Sigma_i]_{\alpha\alpha}\geq 0$ $\forall\,\alpha\leq \nu_i$ are the so-called "singular values", and $[\Sigma_i]_{a\alpha}=0$ $\forall\,a>\nu_i$.
This property arises from the fact that the SVD captures the effective rank of $\R_i$, which is $\nu_i$, and the non-zero singular values correspond to the contributions from the physical $\ca_{i\alpha}$ modes.
In our context of application inherent in the gGA equations, where we know that:
\begin{equation}
\R_i^\dagger \R_i\simeq \mathbf{1}
\label{hyp}
\,,
\end{equation}
is accurately satisfied, we have that $[\Sigma_i]_{a\alpha}\simeq \delta_{a\alpha}$ for $a\leq \nu_i$, as it can be readily verified by noting that:
\begin{align}
    \Sigma_i^\dagger\Sigma_i =V_i^\dagger \R_i^\dagger\R_i V_i\,.
    \label{SdaggerS}
\end{align}

Let us define the gauge transformation $u_i = U_i \bar{V}_i^\dagger$, 
where $\bar{V}_i$ is a unitary block matrix with entries $[\bar{V}_i]_{\alpha\beta}=[{V}_i]_{\alpha\beta}$
$\forall\,\alpha,\beta\leq \nu_i$, $[\bar{V}_i]_{ab}=\delta_{ab}$
$\forall\,a,b\geq \nu_i+1$, and $0$ elsewhere.
By applying such gauge transformation to $\R_i$ we can bring it into a form close to the one corresponding to the ideal scenario of Fig. 1. In fact:
\begin{align}
    \R_i' &= u_i^\dagger \R_i=\bar{V}_i\Sigma_iV_i
    \,,
\end{align}
which has entries: 
\begin{align}
[\R'_i]_{\alpha\beta}&\simeq \delta_{\alpha\beta}\qquad\forall\,\alpha,\beta\leq \nu_i 
\label{Rpstep}
\\
[\R'_i]_{a\alpha}&=0\qquad\forall\,a>\nu_i
\,,
\end{align}
where Eq.~\eqref{Rpstep} holds true because the corresponding block of $\R_i'$ consists of a unitary rotation of the singular values of $\Sigma_i$, which we know to be approximately $1$ 
under our hyphotesys, see Eqs.~\eqref{hyp} and \eqref{SdaggerS}.

In the arguments presented above, it is important to emphasize that the construction allowing us to obtain an $\R_i$ matrix with the ideal form depicted in Fig. 1 is specifically applicable only to the gGA framework, where $B > 1$, as our key hypothesis that $\R_i^\dagger \R_i \simeq \mathbf{1}$ is not satisfied in the case of the classic GA with $B=1$.

In the following subsection, we will provide a concrete illustration of the gauge-fixing construction explained above. Specifically, we will apply these concepts to the simple case of the Hubbard model at particle-hole symmetry, within the gGA framework with $B=3$, as previously studied in Ref.~\cite{Ghost-GA}. This example will serve to further clarify the gauge-fixing procedure and demonstrate the practical utility of the auxiliary ghost modes in capturing the many-body effects arising in this system.

\subsection*{QE gauge fixing for single-band Hubbard model}

Let us consider the special case of a single-band Hubbard model ($\nu_i=1$) satisfying particle-hole symmetry and
translational invariance. The translational invariance
implies that $\R_i=\R$ and $\Lambda_i=\Lambda$ are independent of the fragment $i$.
Furthermore, as previously pointed out in Ref.~\cite{Ghost-GA}, the particle-hole symmetry condition implies that, in the gauge that diagonalizes $\Lambda$, we have:
\begin{align}
    \R=\begin{pmatrix}
    \sqrt{z}   \\
    \sqrt{{h}/{2}} \\
    \sqrt{{h}/{2}}
  \end{pmatrix}
  \,,
  \qquad
    \Lambda=\begin{pmatrix}
    0 & 0    & 0 \\
    0 & l    & 0 \\
    0 & 0    & -l
  \end{pmatrix}
  \,;
\end{align}
where $z$ representes the quasi-particle weight, $h$ represents the spectral weight of the Hubbard bands,
and $l$ controls the position of the Hubbard bands.

In such simple case, the SVD of $\R$:
\begin{align}
    \R&=U\Sigma V^\dagger
\end{align}
can be realized as follows:
\begin{align}
    U&=\frac{1}{\sqrt{z+h}}\begin{pmatrix}
    \sqrt{z} & -\sqrt{h}  & 0 \\
    \sqrt{\frac{h}{2}} & \sqrt{\frac{z}{2}}  &  -\sqrt{\frac{z+h}{2}} \\
    \sqrt{\frac{h}{2}} & \sqrt{\frac{z}{2}}  &  \sqrt{\frac{z+h}{2}}
  \end{pmatrix}
  \\
  \Sigma&=\begin{pmatrix}
    \sqrt{z+h}   \\
    0 \\
    0
  \end{pmatrix}
  \\
  V&=1
  \,,
\end{align}
and the QE gauge fixing procedure above reduces to performing the following matrix multiplications:
\begin{align}
    \R'&=U^\dagger \R=\begin{pmatrix}
    \sqrt{z+h}   \\
    0 \\
    0
  \end{pmatrix}
  \simeq\begin{pmatrix}
    1   \\
    0 \\
    0
  \end{pmatrix}
  \label{Rprime}
    \\
    \Lambda'&=U^\dagger \Lambda U
    =\frac{-l}{\sqrt{z+h}}
    \begin{pmatrix}
    0 & 0    & \sqrt{h} \\
    0 & 0    & \sqrt{z} \\
    \sqrt{h} & \sqrt{z}  &  0
  \end{pmatrix}
  \nonumber\\
  & 
  \simeq -l\begin{pmatrix}
    0 & 0    & \sqrt{h} \\
    0 & 0    & \sqrt{z} \\
    \sqrt{h} & \sqrt{z}  &  0
  \end{pmatrix}
  \label{Lprime}
    \,,
\end{align}
where in the last step of Eqs.~\eqref{Rprime} and \eqref{Lprime} we used that $z+h=\Sigma^\dagger\Sigma\simeq 1$, consistently with the fact that the gGA framework captures both the quasiparticle weight and the Hubbard bands.

Therefore, the gauge transformation above reproduces the scenario represented in Fig.~\ref{Figure1}, featuring the same non-local terms as the original Hamiltonian and additional ghost Fermionic degrees of freedom interacting locally with each fragment. It is important to remember that, as previously explained below Eq.~\eqref{cfR}, this reduction is exact only under the hypothesis that $\R^\dagger \R=z+h=1$, which, while being satisfied accurately in gGA, is not met exactly.

As expected, the auxiliary degrees of freedom (corresponding to the first component of the matrices above) are decoupled from the auxiliary modes in the uncorrelated regime, where the spectral weight $h$ of the Hubbard bands vanishes. On the other hand, such coupling becomes stronger as the interaction strength grows and, in the Mott phase, where the quasiparticle weight $z$ vanishes, it becomes commensurate to $l\simeq \mathcal{U}$, where $\mathcal{U}$ is the Hubbard interaction strength of the model. This is consistent with the notion that $\hat{H}_*$, parametrized by $\R$ and $\Lambda$, serves as an effective Hamiltonian aiming to approximate the many-body interactions between the fragments and their respective environments with one-body operators. Consequently, it is natural that the ghost modes are useful only when many-body interactions are present.

\section{Conclusions}

In this study we derived the
gDMET framework: a QE method based on principles that echo the foundational concepts underlying DMET, that is mathematically equivalent to the gGA.
This reformulation offers an alternative, mathematically precise interpretation of the gGA equations, uncovers the underlying physical link between the two methods, 
and introduces a practical advantage with a simpler implementation. 

A key aspect of our analysis is the crucial role of the ghost degrees of freedom in connecting the DMET and gGA perspectives, i.e., their necessity for formulating a unified framework that adheres to the principles of both approaches.
In relation to this point, it is interesting to note that, as shown in Ref.~\cite{ALM_g-GA}, the limitations of standard GA in capturing charge fluctuations in the strongly correlated regime, particularly in the Mott phase, are directly tied to the method's inability to capture the entire spectral weight, and that the introduction of ghost degrees of freedom in gGA resolves these issues, providing results with accuracy essentially equal to DMFT.
Our study suggests a possible association between the requirement of ghost degrees of freedom for achieving satisfactory accuracy and the feasibility of formulating the gGA equations also from a DMET perspective, contingent on the inclusion of ghost degrees of freedom. 

Finally, the connection between gGA and DMET established here may pave the way for novel methodological generalizations, leveraging on the combined strengths of the variational perspective underlying the gGA framework and the QE perspective underlying DMET.

\section{Acknowledgements}

%
This work was supported by a grant from the Simons Foundation (1030691, NL), and by the the Novo Nordisk Foundation through the Exploratory Interdisciplinary Synergy Programme project
NNF19OC0057790.


%

\newaliascnt{suppeqn}{equation}
\let\oldtheequation\theequation
\let\theequation\thesuppeqn

\clearpage


\onecolumngrid
\begin{center}
\textbf{\large Supplemental material for:\\ Derivation of the Ghost Gutzwiller Approximation from Quantum Embedding principles: the Ghost Density Matrix Embedding Theory}
\vspace{0.4cm}
\end{center}

\setcounter{equation}{0}
\setcounter{figure}{0}
\setcounter{section}{0}
\setcounter{table}{0}
\setcounter{page}{1}
\makeatletter

In Sec.~I of this supplemental material we derive a standard general result in DMET, concerning the Schmidt decomposition of single-particle wavefunctions in bipartite Fermionic systems. 

In Sec.~II, for completeness, we outline the standard variational derivation of the gGA equations as formulated in Ref.~\cite{SM-ALM_g-GA}, with a modified notation consistent with the main text of this work. 

In Sec.~III we demonstrate explicitly the equivalence of the gDMET equations, derived in the main text from a QE perspective, and the standard gGA equations, as derived in Sec.~II from a variational principle.

In Sec.~IV we write explicitly the gDMET equations in momentum representation, for systems with translational symmetry.

\newpage

\maketitle

\section{Schmidt decomposition of Slater determinants}

In this section we prove the standard theorem concerning the Schmidt decomposition of single-particle wavefunctions in bipartite Fermionic systems, following the derivation of Ref.~\cite{SM-dmet-thesis}.

Let us consider a Fermionic system composed by a fragment $A$ generated by Fermionic modes $a^\dagger_i$, $i=1,..,n_A$ and an environment $B$ generated by Fermionic modes $a^\dagger_i$, $i=n_A+1,..,n=n_A+n_B$. 

Given a single particle wavefunction:
\be
\ket{\Psi_0}=c^\dagger_1\hdots c^\dagger_N\,\ket{0}\,,
\ee
we can express the modes $c^\dagger_p$, $p=1,..,N$ in terms of the $a^\dagger_i$ modes as follows:
\begin{align}
c^\dagger_p=\sum_{i=1}^n D_{ip}a^\dagger_i\,,
\end{align}
where $D$ is a $n\times N$ matrix.

We are going to show that it is possible to write explicitly $\ket{\Psi_0}$ in the following form:
\begin{align}
\ket{\Psi_0}=\sum_{k=1}^{2^{n_A}}\sigma_k\ket{\alpha_k}\otimes\ket{\beta_k}\,,
    \label{S-generic}
\end{align}
where $\ket{\alpha_k}$ belongs to $A$ and $\ket{\beta_k}$ belongs to $B$.


It can be readily verified that the single particle density matrix of $\ket{\Psi_0}$ is given by:
\begin{align}
    \rho_{ij}=\Av{\Psi_0}{a^\dagger_i a_j}=[DD^\dagger]_{ji}\,.
    \label{rhoij}
\end{align}

\subsection{Choice of the $c^\dagger_p$ modes}

The state $\ket{\Psi_0}$ is invariant with respect to any unitary transformation of the modes $c^\dagger_p$. Such change of basis corresponds to performing a transformation:
\begin{align}
    D\rightarrow D\Omega\,,
    \label{gauge-C}
\end{align}
where $\Omega$ is an arbitrary $N\times N$ unitary matrix.

By exploiting such freedom it is possible to reduce $D$ to the following block form:
\begin{align}
    D= \left[
\begin{array}{c|c}
P & 0 \\
\hline
Q & E
\end{array}
\right]
\,,
\end{align}
where $P$ is $n_A\times n_A$.
Once such choice is made it is still possible to exploit the freedom [Eq.~\eqref{gauge-C}] applying a unitary transformation of the form:
\begin{align}
    \Omega= \left[
\begin{array}{c|c}
W & 0 \\
\hline
0 & 0
\end{array}
\right]
\,,
\end{align}
therefore modifying $D$ as follows:
\begin{align}
    \left[
\begin{array}{c|c}
P & 0 \\
\hline
Q & E
\end{array}
\right]
\rightarrow
\left[
\begin{array}{c|c}
PW & 0 \\
\hline
QW & E
\end{array}
\right]
\label{gauge-block}
\,,
\end{align}

Within any of such choices of basis for the $c^\dagger_p$ modes, the single particle density matrix of $\ket{\Psi_0}$ is given by the transpose of the following matrix:
\begin{align}
    DD^\dagger= \left[
\begin{array}{c|c}
PP^\dagger & PQ^\dagger \\
\hline
QP^\dagger & QQ^\dagger+PP^\dagger
\end{array}
\right]
\,.
\end{align}
Note that $PP^\dagger$ is the transpose of the restriction to the $A$ system of the single-particle density matrix $\rho$, which we call $\Delta$:
\begin{align}
    ^t\Delta=PP^\dagger\,.
\end{align}
As such, it is also invariant with respect to any transformation of the form [Eq.~\eqref{gauge-block}] (as the whole matrix $^t\rho=DD^\dagger$).

It is useful to note that, instead, $P^\dagger P$ is not invariant under the transformation $P\rightarrow PW$, but it trasnforms as follows:
\begin{equation}
    P^\dagger P\rightarrow W^\dagger P^\dagger P W\,.
\end{equation}
This freedom can be used for diagonalizing $P^\dagger P$. 
We call $n^0$ the diagonal eigenvalue matrix of $P^\dagger P$, and note that these are the same eigenvalues of $^t\Delta=PP^\dagger$ (and, therefore, of $\Delta$ itself).
As we are going to show below, this basis choice is useful for the purpose of deriving an explicit Schmidt decomposition of $\ket{\Psi_0}$ (see Eq.~\eqref{S-generic}). 
Therefore, from now on we will choose to express $\ket{\Psi_0}$ in terms of modes $\cc_p$ such that $P^\dagger P=n^0$ is diagonal.

We observe that, since $D^\dagger D=1$, we have:
\begin{align}
    P^\dagger P + Q^\dagger Q &= 1
    \label{orthPQ}
    \\
    Q^\dagger E &=1
    \label{orthQE}
    \,.
\end{align}
From Eq.~\eqref{orthPQ} it follows that, within our basis choice for the $\cc_p$ modes:
\begin{equation}
    Q^\dagger Q=\mathbf{1}-n^0\,,
\end{equation}
which is also diagonal.

We consider the following matrices:
\begin{align}
    \tilde{P}&=P\frac{1}{\sqrt{P^\dagger P}}=P\frac{1}{\sqrt{n^0}}
    \\
    \tilde{Q}&=Q\frac{1}{\sqrt{Q^\dagger Q}}=Q\frac{1}{\sqrt{\mathbf{1}-n^0}}
    \,,
\end{align}
which satisfy the following equations:
\begin{align}
    \tilde{P}^\dagger \tilde{P}&=1
    \label{pp}
    \\
    \tilde{Q}^\dagger \tilde{Q}&=1
    \label{qq}
    \,,
\end{align}
and define the following operators:
\begin{align}
    \cc_{A,k}&=\sum_{i=1}^{n_A} \tilde{P}_{ik}\, a^\dagger_i\;\quad (k=1,..,n_A)
    \label{An}
    \\
    \cc_{B,k}&=\sum_{j=1}^{n_B} \tilde{Q}_{jk}\, a^\dagger_{j+n_A}\;\quad (k=1,..,n_A)
    \label{Bn}
    \\
    \cc_{B,l}&=\sum_{j=1}^{n_B} E_{jl}\, a^\dagger_{j+n_A}\;\quad (l=n_A+1,..,N)
    \label{C}
    \,,
\end{align}
which, because of Eqs.~\eqref{pp} and \eqref{qq}, are independent modes obeying the canonical anticommutation rules.

Using the definitions above, it can be readily verified that:
\begin{align}
    \ket{\Psi_0}=
    \sum_{\Gamma=0}^{2^{n_A}-1}
    &\prod_{k=1}^{n_A}\left[\sqrt{n^0_{kk}}\right]^{q_k(\Gamma)} \left[\sqrt{1-n^0_{kk}}\right]^{1-q_k(\Gamma)}
    \nonumber\\&
    \left(\ket{A,\Gamma}\otimes\ket{B,\Gamma}\right)\otimes \ket{\psi_C}
    \,,
    \label{Snat}
\end{align}
where:
\begin{align}
    \ket{A,\Gamma}&=[\cc_{A,1}]^{q_1(\Gamma)}\hdots[\cc_{A,n_A}]^{q_{n_A}(\Gamma)}\,\ket{0}
    \\
    \ket{B,\Gamma}&=[\cc_{B,1}]^{1-q_1(\Gamma)}\hdots[\cc_{B,n_A}]^{1-q_{n_A}(\Gamma)}\,\ket{0}
    \\
    \ket{\psi_C}&=\cc_{B,n_A+1}\hdots\cc_{B,N}\,\ket{0}
    \label{SM-psiC}
    \,,
\end{align}
providing us with an explicit realization of Eq.~\eqref{S-generic}.

For later convenience, we calculate the components of the single-particle density matrix in terms of the modes [Eqs.~\eqref{An}-\eqref{C}]:
\begin{align}
    &\Av{\Psi_0}{\cc_{A,k}\ca_{A,k'}}=
    \sum_{i,i'=1}^{n_A}
    \tilde{P}_{ik}\tilde{P}^\dagger_{k'i'}\rho_{ii'}
    \nonumber
    \\
    &\;=\left[
    \frac{1}{\sqrt{P^\dagger P}}P^\dagger PP^\dagger P \frac{1}{\sqrt{P^\dagger P}}
    \right]_{k'k}
    \nonumber
    \\
    &\;=\left[P^\dagger P\right]_{k'k}=\delta_{kk'}n^0_{kk}
    \label{dmnAA}
    \,.
\end{align}
Similarly, we see that:
\begin{align}
    &\Av{\Psi_0}{\cc_{A,k}\ca_{B,k'}}=
    \sum_{i,i'=1}^{n_A}
    \tilde{P}_{ik}\tilde{Q}^\dagger_{k'i'}\rho_{i,i'+n_A}
    \nonumber
    \\
    &\;=\sum_{i,i'=1}^{n_A}
    \tilde{P}_{ik}\tilde{Q}^\dagger_{k'i'}[DD^\dagger]_{i'+n_A,i}
    \nonumber
    \\
    &\;=\sum_{i,i'=1}^{n_A}
    \tilde{P}_{ik}\tilde{Q}^\dagger_{k'i'}[QP^\dagger]_{i'i}
    \nonumber
    \\
    &\;=\left[
    \frac{1}{\sqrt{Q^\dagger Q}}Q^\dagger QP^\dagger P \frac{1}{\sqrt{P^\dagger P}}
    \right]_{k'k}\!\!= \left[P^\dagger P\right]_{k'k}
    \nonumber
    \\
    &\;=\left[
    {\sqrt{Q^\dagger Q}}{\sqrt{P^\dagger P}}
    \right]_{k'k}
    =\delta_{kk'}\sqrt{n^0_{kk}(1-n^0_{kk})}
    \label{dmnAB}
    \,,
\end{align}
and that:
\begin{align}
    &\Av{\Psi_0}{\cc_{B,k}\ca_{B,k'}}=
    \sum_{i,i'=1}^{n_A}
    \tilde{Q}_{ik}\tilde{Q}^\dagger_{k'i'}\rho_{i+n_A,i'+n_A}
    \nonumber
    \\
    &\;=\left[
    \frac{1}{\sqrt{Q^\dagger Q}}Q^\dagger QQ^\dagger Q \frac{1}{\sqrt{Q^\dagger Q}}
    \right]_{k'k}
    \nonumber
    \\
    &\;=\left[Q^\dagger Q\right]_{k'k}=\delta_{kk'}(1-n^0_{kk})
    \label{dmnBB}
    \,.
\end{align}

\subsection{Reformulation in the original basis}

Note that Eq.~\eqref{Snat} depends on $\tilde{P}$
and $\tilde{Q}$, that are obtained from the implicit procedure above.
Below we derive a more practical expression, explicitly written as a function of the single-particle density matrix $\rho$ of $\ket{\Psi_0}$ in the original basis, see Eq.~\eqref{rhoij}.
To achieve this we note that:
\begin{align}
    [DD^\dagger]_{AA}&=^t\Delta=PP^\dagger = \tilde{P} n^0 \tilde{P}^\dagger
    \\
    [DD^\dagger]_{AB}&=PQ^\dagger = \tilde{P} \sqrt{n^0(\mathbf{1}-n^0)}\tilde{Q}^\dagger
    \nonumber\\
    &\qquad
    =\tilde{P} \sqrt{n^0(\mathbf{1}-n^0)}\tilde{P}^\dagger \tilde{P}\,\tilde{Q}^\dagger
    \nonumber\\
    &\qquad
    = \sqrt{^t\Delta(\mathbf{1}-^t\Delta)}\, \tilde{P}\tilde{Q}^\dagger
    \,.
\end{align}
From this we can deduce that:
\begin{equation}
    \tilde{Q}=\, [DD^\dagger]_{BA}\frac{1}{\sqrt{^t\Delta(\mathbf{1}-^t\Delta)}}\,\tilde{P}
    \,.
\end{equation}

By exploiting the unitarity of $\tilde{P}$ and multiplying Eqs.~\eqref{An} and \eqref{Bn} by $\tilde{P}^\dagger$ on the right hand side, we obtain the following set of independent anticommuting Fermionic modes:
\begin{align}
    &a^\dagger_k\;\quad (k=1,..,n_A)
    \label{an}
    \\
    b^\dagger_{k}&=\sum_{j=1}^{n_B} \B_{jk}\, a^\dagger_{j+n_A}\;\quad (k=1,..,n_A)
    \label{bn}
    \\
    c^\dagger_{B,l}&=\sum_{j=1}^{n_B} E_{jl}\, a^\dagger_{j+n_A}\;\quad (l=n_A+1,..,N)
    \label{Crep}
    \,,
\end{align}
where:
\begin{align}
    \B&=\tilde{Q}\tilde{P}^\dagger =\, [DD^\dagger]_{BA}\frac{1}{\sqrt{^t\Delta(\mathbf{1}-^t\Delta)}}
    \,.
\end{align}

From the equations above it follows that, $\forall\,j=1,..,n_B$:
\begin{align}
    a^\dagger_{j+n_A}=\sum_{k=1}^{n_A} \B^\dagger_{kj}\,b^\dagger_k
    + \hdots
    \,,
\end{align}
where the dots indicate a linear combination of the fully-occupied modes $\cc_{B,l}$ ($l=n_A+1,..,N$)
and the remaining fully-empty modes in $\ket{\Psi_0}$.

\subsection{Quantum embedding (QE) projection}

Let us consider a single particle Hamiltonian lying within the Hilbert space spanned by the 
Fermionic modes $a^\dagger_i$, $i=1,..,n_A$ and $a^\dagger_i$, $i=n_A+1,..,n=n_A+n_B$:
\begin{align}
    H&=\sum_{i,j=1}^{n} h_{ij}\,a^\dagger_i a_j
    \nonumber\\
    &=\sum_{i,j=1}^{n_A}h^{AA}_{ij}\,a^\dagger_i a_j+ 
    \sum_{i,j=1}^{n_B}h^{BB}_{ij}\,a^\dagger_{i+n_A} a_{j+n_A}
    \nonumber\\
    &+\sum_{i=1}^{n_A}\sum_{j=1}^{n_B}\left[
    h^{AB}_{ij}\,a^\dagger_i a_{j+n_A}+\text{H.c.}
    \right]\,.
\end{align}

From the Schmidt decomposition of the ground state $\ket{\Psi_0}$ of $H$, we construct the embedding projector:
\begin{align}
\P_A=
\sum_{\Gamma,\Gamma'}
\delta_{N_\Gamma+N_{\Gamma'},n_A}
\left(\ket{A,\Gamma}\!\otimes\!\ket{B,\Gamma'}\right)
\left(\bra{A,\Gamma}\!\otimes\!\bra{B,\Gamma'}\right)
\otimes \P^A_C
\,,
\end{align}
where:
\begin{equation}
    \P^A_C=\ket{\psi_C}\bra{\psi_C}
    \,,
\end{equation}
$N_{\Gamma}$ indicates the number of Fermions in the multiplet $\Gamma$.
The $\delta_{N_\Gamma+N_{\Gamma'},n_A}$
encodes the fact that, since $\ket{\Psi_0}$ has $N$ Fermions
 and $
\ket{\Psi_C}$ has $N-n_A$ Fermions (see Eq.~\eqref{SM-psiC}),
the $\ket{A,\Gamma}$ and $\ket{B,\Gamma}$ multiplets combined
contain $N-(N-n_A)=n_A$ Fermions.

Note that, by construction the projector $\P_A$ satisfies the identity:
\begin{equation}
    \P_A\ket{\Psi_0}=\ket{\Psi_0}\,.
\end{equation}

It can be straightforwardly verified that:
\begin{align}
    H^{emb}_A&=\P_A H \P_A 
    \\
    &=\P_A\left[\sum_{i,j=1}^{n_A}h^{AA}_{ij}\,a^\dagger_i a_j+ \!\!
    \sum_{q,q'=1}^{n_B}\!\!\left[\B^\dagger h^{BB} \B\right]_{qq'}b^\dagger_q b_{q'}\right.
    \nonumber\\
    &\left.+\sum_{i,q=1}^{n_A}\left(
    \left[h^{AB}\B\right]_{iq}\,a^\dagger_i b_q+\text{H.c.}
    \right)+\text{const}\right]\P_A
    \nonumber
    \,.
\end{align}

The single-particle density matrix for the $a_i$ and $b_j$ modes can be computed, e.g., 
from Eqs.~\eqref{dmnAA}-\eqref{dmnBB} and the fact that:
\begin{align}
    a^\dagger_i&=\sum_{k=1}^{n_A}\tilde{P}_{ki}\cc_{A,k}
    \\
    b^\dagger_i&=\sum_{k=1}^{n_A}\tilde{P}_{ki}\cc_{B,k}
    \,.
\end{align}
From these equations one readily obtains that:
\begin{align}
\Av{\Psi_0}{a^\dagger_i a_j}&=\Delta_{ij}
\\
\Av{\Psi_0}{a^\dagger_i b_j}&=[\Delta(\mathbf{1}-\Delta)]^{\frac{1}{2}}_{ij}
\\
\Av{\Psi_0}{b^\dagger_i b_j}&=[\mathbf{1}-\Delta]_{ij}
\,.
\end{align}

\subsection{Summary}

In summary, given a generic quadratic Fermionic Hamiltonian:
\begin{align}
    H&=\sum_{i,j=1}^{n} h_{ij}\,a^\dagger_i a_j
    \nonumber\\
    &=\sum_{i,j=1}^{n_A}h^{AA}_{ij}\,a^\dagger_i a_j+ 
    \sum_{i,j=1}^{n_B}h^{BB}_{ij}\,a^\dagger_{i+n_A} a_{j+n_A}
    \nonumber\\
    &+\sum_{i=1}^{n_A}\sum_{j=1}^{n_B}\left[
    h^{AB}_{ij}\,a^\dagger_i a_{j+n_A}+\text{H.c.}
    \right]\,.
\end{align}
where $n=n_A+n_B$,
we deduced that $\ket{\Psi_0}$ is the tensor product of the ground state of
\begin{align}
H^{emb}_A&=\sum_{i,j=1}^{n_A}h^{AA}_{ij}\,a^\dagger_i a_j+ \!\!
    \sum_{q,q'=1}^{n_A}\!\!\left[\B^\dagger h^{BB} \B\right]_{qq'}b^\dagger_q b_{q'}
    \nonumber\\
    &+\sum_{i,q=1}^{n_A}\left(
    \left[h^{AB}\B\right]_{iq}\,a^\dagger_i b_q+\text{H.c.}
    \right)
\end{align}
within the subspace with half-filling (i.e., with $n_A$ Fermions), and a state 
 $\ket{\Psi_C}$,
where:
\begin{align}
    b^\dagger_{k}&=\sum_{j=1}^{n_B} \B_{jk}\, a^\dagger_{j+n_A}\;\quad (k=1,..,n_A)
    \label{SM-Bn-recap}
    \,,
\end{align}
the entries of the matrix $\B$ are given by the following equation:
\begin{align}
    \B_{jk} &= \sum_{l=1}^{n_A}
    \left[\frac{1}{\sqrt{\Delta(\mathbf{1}-\Delta)}}\right]_{kl}
    \Av{\Psi_0}{a^\dagger_la_j}
    \\
    &= \sum_{l=1}^{n_A}
    \left[\frac{1}{\sqrt{\Delta(\mathbf{1}-\Delta)}}\right]_{kl}
    [f(h)]_{jl}
    \\
    &= 
    \left[
    f(h)\frac{1}{\sqrt{^t\Delta(\mathbf{1}-^t\Delta)}}
    \right]_{jk}
    \,,
\end{align}
where $f$ is the Fermi function, $l=1,..,n_A$, $j=1,..,n_B$, 
\begin{align}
    \Delta_{ij}&=\Av{\Psi_0}{a^\dagger_i a_j}\;\quad(i,j=1,..,n_A)
    \nonumber
    \\
    &=[\Pi_Af(h)\Pi_A]_{ji}
    \,.
\end{align}



The other entries of the single particle density matrix of $H^{emb}_A$ are the following:
\begin{align}
\Av{\Psi_0}{a^\dagger_i b_j}&=[\Delta(\mathbf{1}-\Delta)]^{\frac{1}{2}}_{ij}
\\
\Av{\Psi_0}{b^\dagger_i b_j}&=[\mathbf{1}-\Delta]_{ij}
\,.
\end{align}

The state $\ket{\Psi_C}$ is generated by other independent modes (that can be calculated, but we are are not going to need explicitly).

The physical meaning of the result above is that, given a quadratic Hamiltonian and a fragment $A$,
the ground state $\ket{\Psi_0}$ can be expressed in terms of the $2n_A$ "active" degrees of freedom $a_i$ (impurity) and $b_i$ (bath); while the remaining "core" degrees of freedom, consisting of $n-n_A$ fully occupied states and $n_A+n_B-N$ empty states, are decoupled from the fragment $A$ and its bath $B$.
The equations above provide us with an explicit expression for the many-body active component of $\ket{\Psi_0}$ in terms of its single-particle density matrix.

Within the context of QE theories, discussed in the main text, the result above can be used for constructing self-consistently a state $\ket{\Psi_0}$ identifying the active degrees of freedom of any fragment of an extended correlated system.

\section{Standard variational formulation of the $\text{g}$GA equations}

For completeness, in this section we provide a comprehensive derivation of the gGA method from the variational perspective, as formulated in Ref.~\cite{SM-ALM_g-GA}, but with a notation consistent with the main text of this paper.

Let us consider a generic multi-orbital Fermi-Hubbard Hamiltonian represented as in the main text:
\begin{align}
    \hat{H}&=\sum_{i=1}^{\mathcal{N}}\hat{H}_{loc}^i[\cc_{i\alpha},\ca_{i\alpha}]+\sum_{i\neq j}\hat{T}_{ij}
    \label{SM-H}
    \\
    \hat{T}_{ij}&=\sum_{\alpha=1}^{\nu_i}\sum_{\beta=1}^{\nu_j}[t_{ij}]_{\alpha\beta}\,\cc_{i\alpha}\ca_{j\beta}
    \,,
\end{align}
where $i$ and $j$ label the fragments of the system, $\hat{H}_{loc}^i$ is a generic operator lying within the $i$ fragment (i.e., constructed with $\cc_{i\alpha},\ca_{i\alpha}$), including both one-body and two-body contributions, and $\alpha$ labels all Fermionic modes within each fragment.

\begin{figure} 
    \includegraphics[width=7.7cm]{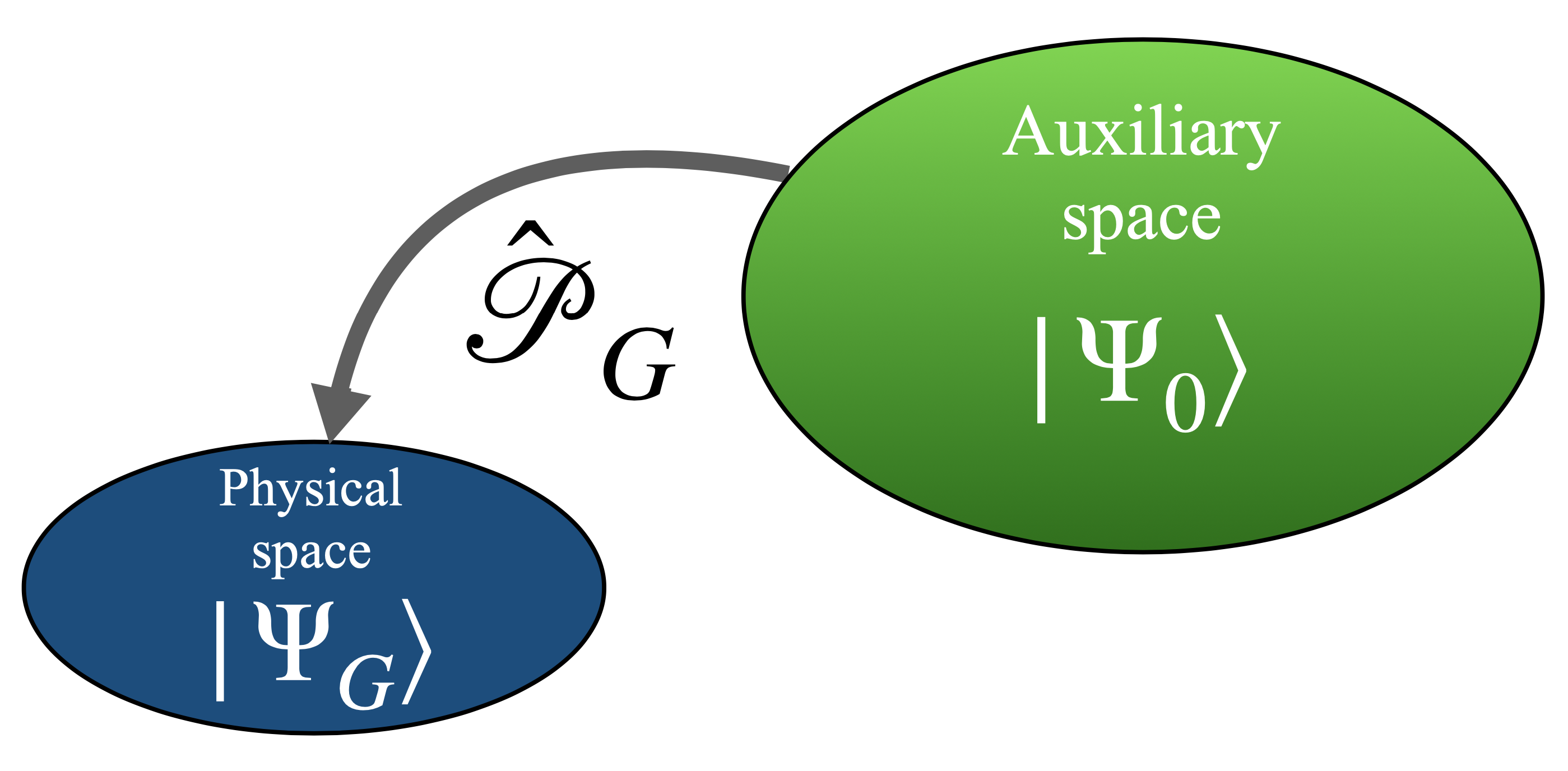}
    \caption{Schematic representation of the gGA variational ansatz. The wavefunction $\ket{\Psi_G}$ is constructed by mapping a generic single-particle wavefunction $\ket{\Psi_0}$ constructed in an auxiliary Hilbert space, using the operator $\proj$.
    Both $\ket{\Psi_0}$ and $\proj$ are determined variationally}
    \label{FigureS1}
\end{figure}

\subsection{The $\text{g}$-GA variational ansatz}

The gGA consists in minimizing the variational energy
with respect to a wave function represented as follows:
\begin{align}
    \ket{\Psi_G}&=\proj\ket{\Psi_0}
\label{GWF}
\\
\proj&=\prod_{i=1}^{\mathcal{N}}\projRi
\,,
\end{align}
where $\ket{\Psi_0}$ is a single-particle wavefunction constructed in an auxiliary Hilbert space, generated by $B{\nu}_i$ degrees of freedom $\fc_{ia}$ $(a=1,..,B{\nu}_i)$ for each $i$,  $B\geq 3$ is an integer odd number, and:
\begin{align}
\projRi&=\sum_{\Gamma=0}^{2^{\nu_i}-1}
\,
\sum_{n=0}^{2^{B{\nu}_i}-1}
[\Lambda_{i}]_{\Gamma n}
\ket{\Gamma,{i}}\bra{n,{i}}
\\
\ket{\Gamma,{i}}&=[\cc_{i1}]^{q_1(\Gamma)} ...
[\cc_{i q_{\nu_i}}]^{q_{\nu_i}(\Gamma)}\,\ket{0}
\\
\ket{n,{ i}}&=[\fc_{i1}]^{q_1(n)} ...
[\fc_{i q_{B{\nu}_i}}]^{q_{B{\nu}_i}(n)}\,\ket{0}
\end{align}
is an operator mapping the local auxiliary-space states into the physical space, see Fig.~\ref{FigureS1}, where $q_i(j)$ represents the $i$-th digit of $j$ in binary representation, and the matrix $\Lambda_i$ controls how $\projRi$
modifies the weight of the local electronic configurations.
The key reason why the gGA variational ansatz generalizes the standard multi-orbital GA theory is that $B{\nu}_i>\nu_i$.

Here we focus on the normal phase. We observe that, for enforcing the symmetry condition that $\ket{\Psi_G}$ is an eigenstate of the number operator, we do \emph{not}
need to assume that $\projRi$ commutes with the number operator, but only that:
\be
\sum_{j=1}^{B{\nu}_i} q_j(n)-
\sum_{j=1}^{\nu_i} q_j(\Gamma)=m_i\quad
\forall\,\Gamma,n \, | \,
[\Lambda_i]_{\Gamma n}\neq 0\,,
\label{mi}
\ee
where $m_i$ is integer.
In principle, $m_i$ could be regarded as an arbitrary variational parameter).
In the present work, following Ref.~\cite{SM-ALM_g-GA}, we assume that $B\geq 3$ is an integer odd number, and set $m_i=(B-1)\nu_i/2$ (see Eq.~\eqref{mi}).
The standard GA theory is recovered in the special case $B=1$.

As in the standard multi-orbital GA, the variational wave function is restricted by the following
conditions:
\begin{align}
\Av{\Psi_0}{\projRidagger\projRi} &= \langle\Psi_0|\Psi_0\rangle=1
\label{gconstr1}
\\
\Av{\Psi_0}{\projRidagger\projRi\,\fc_{ia}\fa_{ib}} &=
\Av{\Psi_0}{\fc_{ia}\fa_{ib}}\qquad\forall\,a,b=1,...,B{\nu}_i 
\,,
\label{gconstr2}
\end{align}
which are commonly called "Gutzwiller constraints".
Furthermore, the so-called "Gutzwiller Approximation", which becomes
exact in the limit of infinite coordination number~\cite{SM-Gutzwiller3,SM-GA-infinite-dim}
---where Dynamical Mean Field Theory (DMFT) is exact~\cite{SM-DMFT}--- is assumed.

\subsection{The $\text{g}$-GA variational-energy contributions}

As in the standard multi-orbital GA, our goal is to evaluate the variational energy:
\be
\mathcal{E}(\Psi_0,\proj)=\Av{\Psi_0}{\proj^\dagger\hat{H}\proj^\dagga}\,.
\label{varen}
\ee
The only difference is that $\ket{\Psi_0}$ resides within the auxiliary (extended) Hilbert space generated by the $\fc_{Ria}$ operators, where $a\in\{1,..,B{\nu}_i\}$ and $B{\nu}_i\geq \nu_i$.

By employing the Gutzwiller approximation~\cite{SM-Gutzwiller3,SM-GA-infinite-dim}
and enforcing the Gutzwiller constraints [Eqs.~\eqref{gconstr1},\eqref{gconstr2}], it can be shown that:
\begin{align}
\Av{\Psi_0}{\proj^\dagger\,\cc_{i\alpha}\ca_{j\beta}\,\proj}
&=\sum_{a=1}^{B{\nu}_i}\sum_{b=1}^{B{\nu}_j}
\Av{\Psi_0}{\left([\mathcal{R}_i]_{a\alpha}\fc_{ia}\right)\left([\mathcal{R}_j]^\dagger_{\beta b}\fa_{jb}\right)}
\qquad\forall\,i\neq j\,,
\label{hoppav}
\\
\Av{\Psi_0}{\proj^\dagger\,\hat{H}_{loc}^i[\cc_{i\alpha},\ca_{i\alpha}]\,\proj}
&=\Av{\Psi_0}{\projRidagger\,\hat{H}_{loc}^i[\cc_{i\alpha},\ca_{i\alpha}]\,\projRi}\,,
\label{locav}
\end{align}
where the $B{\nu}_i\times \nu_i$ matrices $\mathcal{R}_i$ are the solution of the following linear equation:
\be
\Av{\Psi_0}{\projRidagger \cc_{i\alpha}\projRi \fa_{ia}}
=\sum_{b=1}^{B{\nu}_i}[\mathcal{R}_i]_{b\alpha}
\Av{\Psi_0}{\fc_{i b} \fa_{ia}}\,.
\label{Rdef}
\ee

\subsection{Expectation values of local observables with respect to $\ket{\Psi_0}$}

The evaluation of the total-energy components [Eqs.~\eqref{hoppav}-\eqref{Rdef}] and the Gutzwiller constraints [Eqs.~\eqref{gconstr1},\eqref{gconstr2}] all involve expectation values with respect to $\ket{\Psi_0}$ of "local operators" (i.e., involving only $\fc_{ia}$ and $\fa_{ia}$ degrees of freedom at fixed $i$).
To evaluate these quantities,
it is useful to introduce the so-called "local reduced density matrix" of $\ket{\Psi_0}$.

By exploiting the fact that Wick's theorem applies to $\ket{\Psi_0}$ (since it is a single-particle wavefunction), it can be readily verified that its reduced density matrix to the $i$ subsystem is given by:
\be
\hat{P}^0_{i}\propto \exp\left\{-\sum_{a,b=1}^{B{\nu}_i} \left[\ln\left(\frac{\mathbf{1}-{^t\!\Delta_i}}{^t\!\Delta_i}\right)\right]_{ab}
\fc_{i a} \fa_{ib}
\right\}\,,
\label{hatp0}
\ee
where the entries of the $B{\nu}_i\times B{\nu}_i$ matrix $\Delta_i$ are given by:
\be
[\Delta_i]_{ab}=\Av{\Psi_0}{\fc_{i a} \fa_{ib}}
\ee
and ${^t\!\Delta_i}$ indicates the transpose of $\Delta_i$.

From the definitions above, it can be straightforwardly verified that:
\begin{align}
    \Av{\Psi_0}{\projRidagger\projRi}&=
    \Tr\big[P^0_i\Lambda^\dagger_i\Lambda^\dagga_i
    \big]
    \label{loc1}
    \\
    \Av{\Psi_0}{\projRidagger\projRi\,\fc_{ia}\fa_{ib}}&=
    \Tr\big[P^0_i\Lambda^\dagger_i\Lambda^\dagga_i
    \tilde{F}_{ia}^\dagger \tilde{F}_{ib}^\dagga
    \big]
    \\
    \Av{\Psi_0}{\projRidagger\,\hat{H}_{loc}^i[\cc_{i\alpha},\ca_{i\alpha}]\,\projRi}&=
    \Tr\big[P^0_i\Lambda^\dagger_i
    \hat{H}_{loc}^i[F^\dagger_{i\alpha},F^\dagga_{i\alpha}]
    \Lambda^\dagga_i\big]
    \\
    \Av{\Psi_0}{\projRidagger \cc_{i\alpha}\projRi \fa_{ia}}&=
    \Tr\big[P^0_i\Lambda^\dagger_i
    F_{i\alpha}^\dagger \Lambda^\dagga_i \tilde{F}_{ib}^\dagga\big]
    \label{loc4}
    \,,
\end{align}
where $\Tr$ is the trace and:
\begin{align}
    [{P}^0_{i}]_{nn'}&=\langle n,i |
    \hat{P}^0_{i}
    | n', i\rangle
    \qquad (n,n'\in\{0,..,2^{B{\nu}_i}-1\})
    \\
    [F_{i\alpha}]_{\Gamma\Gamma'}&=
    \langle \Gamma,i |
    \ca_{i\alpha}
    | \Gamma', i\rangle
    \qquad (\Gamma,\Gamma'\in\{0,..,2^{{\nu}_i}-1\})
    \label{FGamma}
    \\
    [\tilde{F}_{ia}]_{nn'}&=
    \langle n,i |
    \fa_{i a}
    | n', i\rangle
    \qquad (n,n'\in\{0,..,2^{B{\nu}_i}-1\})
    \label{Fn}
    \,.
\end{align}

\subsection{Matrix of slave-boson amplitudes}

Following Refs.~\cite{SM-lanata-barone-fabrizio,SM-equivalence_GA-SB,SM-Our-PRX}, 
we introduce the so-called matrix of slave-boson (SB)
amplitudes~\cite{SM-Fresard1992,SM-Georges-RISB,SM-Lanata2016}:
\be
\phi_i=\Lambda_i\sqrt{P^0_i}\,.
\label{sba}
\ee

By substituting Eq.~\eqref{sba} in Eqs.~\eqref{loc1}-\eqref{loc4}, it can be readily verified that the Gutzwiller constraints can be rewritten as follows:
\begin{align}
\Tr\big[\phi^\dagger_i\phi^\dagga_i\big] &= \langle\Psi_0|\Psi_0\rangle=1
\label{gconstr1sb}
\\
\Tr\big[\phi^\dagger_i\phi^\dagga_i \tilde{F}_{ia}^\dagger \tilde{F}_{ib}^\dagga\big]
    &=
\Av{\Psi_0}{\fc_{ia}\fa_{ib}}=[\Delta_i]_{ab}
\qquad\forall\,a,b=1,...,B{\nu}_i 
\label{gconstr2sb}
\end{align}
and that Eq.~\eqref{Rdef} can be rewritten as follows:
\be
\Tr\big[\phi^\dagger_i
    F_{i\alpha}^\dagger \phi^\dagga_i \tilde{F}_{ia}^\dagga\big]
=\sum_{c=1}^{B{\nu}_i}[\mathcal{R}_i]_{c\alpha}\,
[\Delta_i(\mathbf{1}-\Delta_i)]^{\frac{1}{2}}_{ca}
\,.
\label{Rdefsb}
\ee

\subsection{Embedding mapping}

Following Ref.~\cite{SM-Our-PRX}, we introduce the so called "embedding states," which are related to the SB amplitudes as follows:
\begin{align}
    \ket{\Phi_i}=\sum_{\Gamma=0}^{2^{\nu_i}-1} \,\sum_{n=0}^{2^{B{\nu}_i}-1}
    e^{i\frac{\pi}{2}N(n)(N(n)-1)}
    \,[\phi_i]_{\Gamma n} \,
    \ket{\Gamma;i}\otimes U_{\text{PH}}\ket{n;i}
    \label{ehstates}
\end{align}
where
\begin{align}
    \ket{\Gamma;i}&= [\cc_{i1}]^{q_1(\Gamma)} ...
[\cc_{i {B{\nu}_i}}]^{q_{{\nu}_i}(\Gamma)}\,\ket{0}
    \\
    \ket{n;i}&= [\bbc_{i1}]^{q_1(n)} ...
[\bbc_{i {B{\nu}_i}}]^{q_{B{\nu}_i}(n)}\,\ket{0}
    \,,
\end{align}
$U_{\text{PH}}$ is a particle-hole transformation acting over the $\ket{n;i}$ states and
\be
N(n)=\sum_{a=1}^{B{\nu}_i} q_a(n)
\,.
\ee

Note that the set of all embedding states represented as in Eq.~\eqref{ehstates} constitute a Fock space, corresponding to an "impurity" (generated by the Fermionic degrees of freedom $\ca_{i\alpha}$, $\alpha\in\{1,..,\nu_i\}$) and a "bath" (generated by the Fermionic degrees of freedom $\bba_{i a}$, $a\in\{1,..,B{\nu}_i\}$).

From the definitions [Eqs.~\eqref{sba},\eqref{ehstates}] and the fact that we assumed $m_i=(B-1)\nu_i/2$ (see Eq.~\eqref{mi}), it follows that this is equivalent to assume that 
$\ket{\Phi_i}$ has a total of $(B+1){\nu}_i/2$ electrons (i.e., that the embedding states are half-filled).

It can be readily verified by inspection that the Gutzwiller constraints can be rewritten as follows:
\begin{align}
\langle\Phi_i|\Phi_i\rangle &= \langle\Psi_0|\Psi_0\rangle=1
\label{gconstr1eh}
\\
\Av{\Phi_i}{\bba_{i b}\bbc_{i a}}
&=
\Av{\Psi_0}{\fc_{ia}\fa_{ib}}=[\Delta_i]_{ab}
\qquad\forall\,a,b=1,...,B{\nu}_i 
\,,
\label{gconstr2eh}
\end{align}
the expectation value of the local terms of $\hat{H}$ can be calculated as:
\be
\Av{\Psi_0}{\projRidagger\,\hat{H}_{loc}^i[\cc_{i\alpha},\ca_{i\alpha}]\,\projRi} =
\Av{\Phi_i}{\hat{H}_{loc}^i[\cc_{i\alpha},\ca_{i\alpha}]}
\ee
and that Eq.~\eqref{Rdef} can be rewritten as:
\be
\Av{\Psi_0}{\cc_{i \alpha}\bba_{i a}}
=\sum_{c=1}^{B{\nu}_i}[\mathcal{R}_i]_{c\alpha}\,
[\Delta_i(\mathbf{1}-\Delta_i)]^{\frac{1}{2}}_{ca}
\,.
\label{Rdefeh}
\ee

In summary, by substituting Eqs.~\eqref{hoppav} and \eqref{locav} in Eq.~\eqref{varen}
and using the equations above,
we deduce that the variational energy can be expressed as follows:
\be
\mathcal{E}=
\sum_{i,j=1}^{\mathcal{N}}\sum_{a,b=1}^{B{\nu}_i}\left[\R_i^\dagga t_{ij}\R^{\dagger}_j\right]_{ab} \fc_{ia}\fa_{jb}
+
\sum_{i= 1}^{\mathcal{N}}
\Av{\Phi_i}{\hat{H}_{loc}^i[\cc_{ i\alpha},\ca_{i\alpha}]}
\,,
\ee
which has to be minimized with respect to the variational parameters
while fulfilling the Gutzwiller constraints [Eqs.~\eqref{gconstr1eh}, \eqref{gconstr2eh}].

\subsection{Lagrange formulation of the gGA} \label{covariance-sec}

As explained in Refs.~\onlinecite{SM-Our-PRX,SM-Lanata2016,SM-Ghost-GA}, the gGA variational-optimization problem formulated above can be solved by extremizing the following Lagrange function:

\bea
&&\Lag[
{\Phi},E^c;\,  \R,\Lambda;\, \D, \Lambda^{c};\,\Delta, \Psi_0, E]=
\Av{\Psi_0}{\h_{\text{qp}}[\R,\Lambda]}
+E\!\left(1\!-\!\langle\Psi_0|\Psi_0\rangle\right)
\nonumber\\&&\quad\quad
+\sum_{i=1}^{\mathcal{N}}\left[\Av{\Phi_i}{\h_i^{\text{emb}}[\D_i,\Lambda_i^c]}
+E^c_i\!\left(\mathbf{1}-\langle \Phi_i | \Phi_i \rangle
\right)\right]
\nonumber\\&&\quad\quad
-\sum_{i=1}^{\mathcal{N}}\left[
\sum_{a,b=1}^{B{\nu}_i}\big(
\left[\Lambda_i\right]_{ab}+\left[\Lambda^c_i\right]_{ab}\big)\left[\Delta_{i}\right]_{ab}
+\sum_{c,a=1}^{B{\nu}_i}\sum_{\alpha=1}^{\nu_i}
\big(
\left[\D_{i}\right]_{a\alpha}\left[\R_{i}\right]_{c\alpha}
\left[\Delta_{i}(\mathbf{1}-\Delta_{i})\right]^{\frac{1}{2}}_{ca}
+\text{c.c.}\big)
\right]\,,
\label{Lag-SB-emb}
\eea
where $\mathcal{N}$ is the total number of unit cells,
$E$ and $E^c$ are real numbers, $\Delta_i$, $\Lambda^c_i$ and $\Lambda_i$ are
$B{\nu}_i\times B{\nu}_i$ Hermitian matrices, $\D_i$ and $\R_i$ are rectangular $B{\nu}_i\times {\nu}_i$ matrices.
The auxiliary Hamiltonians $\h_{\text{qp}}$ and $\h_{\text{emb}}$,
which are called "quasiparticle Hamiltonian" and "Embedding Hamiltonian,"
respectively, are defined as follows:
\begin{align}
    \hat{H}_*[\R,\Lambda]&=\sum_{i,j=1}^{\mathcal{N}}\sum_{a,b=1}^{B{\nu}_i}\left[\R_i^\dagga t_{ij}\R^{\dagger}_j\right]_{ab}
    \fc_{ia}\fa_{jb}
    + \sum_{i=1}^{\mathcal{N}}\sum_{a,b=1}^{B{\nu}_i} \left[ \Lambda_i \right]_{ab}\fc_{ia} \fa_{ib}
    \label{SM-Hqp}
    \\
    \hat{H}^i_{emb}[\D_i,\Lambda^c_i]&= \hat{H}_{loc}^i\big[\cc_{i\alpha},\ca_{i\alpha}\} \big] + 
    \sum_{a=1}^{B{\nu}_i}\sum_{\alpha=1}^{\nu_i}\left(\left[\D_i\right]_{a\alpha}\cc_{i\alpha}\bba_{ia}
    +\text{H.c.}\right)
   +\sum_{a,b=1}^{B{\nu}_i}\left[\Lambda^c_i\right]_{ab}\bba_{ib}\bbc_{ia}
   \,,
   \label{SM-hemb}
\end{align}
which coincide with the definitions emerged in the main text from the gDMET perspective.

As in the main text, for later convenience, we rewrite Eq.~\eqref{SM-Hqp} as follows:
\begin{align}
    \hat{H}_*[\R,\Lambda]
    &=\sum_{i,j=1}^{\mathcal{N}}
    [\Pi_i h_* \Pi_j]_{ab}\,\fc_{ia}\fa_{jb}
    \,,
    \label{SM-Hqp2}
\end{align}
where we introduced the matrix:
\begin{align}
    h_* = \begin{pmatrix}
    \Lambda_1  &\R_1t_{12}\R^\dagger_2& \dots & \R_{1}t_{1 \mathcal{N}}\R^\dagger_{\mathcal{N}} \\
    \R_2t_{21}\R^\dagger_1 & \Lambda_2 & \dots & \vdots \\
    \vdots  & \vdots  & \ddots & \vdots \\
    \R_{\mathcal{N}1}t_{\mathcal{N} 1}\R^\dagger_{1} & \dots& \dots & \Lambda_{\mathcal{N}}
  \end{pmatrix}
  \label{SM-h*}
\end{align}
and the projectors over the degrees of freedom corresponding to each fragment:
\begin{align}
   \label{SM-Proj}
   \Pi_i = \begin{pmatrix}
     \delta_{i1}\left[\mathbf{1}\right]_{B{\nu}_1\times B{\nu}_1} & \dots & \mathbf{0} \\
     \vdots  & \ddots & \vdots \\
    \mathbf{0}  &  \dots & \delta_{iM}\left[\mathbf{1}\right]_{B{\nu}_M\times B{\nu}_M}
  \end{pmatrix}
  \,,
\end{align}
where $\left[\mathbf{1}\right]_{n\times n}$ is the $n\times n$ identity matrix.

The saddle-point of the gGA Lagrange function $\Lag$ defined in Eq.~\eqref{Lag-SB-emb} is given by the following equations:
\begin{align}
    \hat{H}_*[\R,\Lambda]\ket{\Psi_0}&=E_0\ket{\Psi_0}
    \label{SM-Hqp-summary}
    \\
    [\Delta_i]_{ab}&=\Av{\Psi_0}{\fc_{ia} \fa_{ib}}
    \label{SM-Delta-summary}
    \\    \sum_{c,a=1}^{B{\nu}_i}\sum_{\alpha=1}^{\nu_i}\left[\D_i\right]_{c\alpha}\left[\Delta_i\left(\mathbf{1}-\Delta_i\right)\right]^{\tfrac{1}{2}}
    &= 
    \left[
    t_{ij}\R^{\dagger}_j \Pi_jf\left(h_*\right)\Pi_i\right]_{\alpha a} 
    \label{detD}
    \\
    [l^c_i]_{s}
   &=-[l_i]_{s}-\sum_{c,b=1}^{B{\nu}_i}\sum_{\alpha=1}^{\nu_i}\frac{\partial}{\partial \left[d^0_i\right]_s} \left(\left[\Delta_i\left(\mathbf{1}-\Delta_i\right)\right]^{\tfrac{1}{2}}_{cb}\left[\D_i\right]_{b\alpha}\left[\R_i\right]_{c\alpha} + \mathrm{c.c.}\right) 
   \label{detLc}
   \\
   \hat{H}^i_{emb}\ket{\Phi_i} &= E_i^c\ket{\Phi_i}
   \label{SM-Hemb-summary}
   \\
   \bra{\Phi_i}\bba_{ib}\bbc_{ia}\ket{\Phi_i} 
   &=\left[\Delta_i\right]_{ab} 
   \label{detF2}
    \\
    \bra{\Phi_i}\cc_{i\alpha}\bba_{ia}\ket{\Phi_i}
   &= \sum_{c=1}^{B\nu_i}\left[\Delta_i\left(\mathbf{1}-\Delta_i\right)\right]^{\tfrac{1}{2}}\left[\R_i\right]_{c\alpha} 
   \label{detF1}
      \,,
\end{align}
where $f$ is the zero-temperature Fermi function (as in the main text), 
and in Eq.~\eqref{detLc} we introduced the following expansions of the matrices $\Delta_i$, $\Lambda_i$, $\Lambda^c_i$, in terms of an orthonormal
basis of Hermitian matrices $\left\{\left[h_i\right]_s\right\}$ (with respect to the canonical scalar product $(A, B) = \Tr \left[A^{\dagger}B\right]$):
\begin{align}
    \label{coeffDelta}
    \Delta_i =& \sum_{s=1}^{(B{\nu}_i)^2} \left[d^0_i\right]_s {}^t\left[h_i\right]_s \\
    \label{coeffL}
    \Lambda_i =& \sum_{s=1}^{(B{\nu}_i)^2} \left[l_i\right]_s \left[h_i\right]_s  \\
    \label{coeffLc}
    \Lambda^c_i =& \sum_{s=1}^{(B{\nu}_i)^2} \left[l^c_i\right]_s \left[h_i\right]_s \,,
\end{align}
where $\left[d^0_i\right]_s$, $\left[l_i\right]_s$ and $\left[l^c_i\right]_s$ are real-valued coefficients.

\section{Proof of equivalence between the standard $\text{g}$GA equations and the $\text{g}$DMET equations derived in the main text}\label{Lcproof-section}

Here we directly compare the gGA equations [Eqs.~\eqref{SM-Hqp-summary}-\eqref{detF2}], derived above from the standard variational perspective, with the following gDMET equations, derived
in the main from QE principles reminiscent of DMET:

\begin{align}
    \hat{H}_*[\R,\Lambda]\ket{\Psi_0}&=E_0\ket{\Psi_0}
    \label{SM-Hqp-summary2}
    \\
    [\Delta_i]_{ab}&=\Av{\Psi_0}{\fc_{ia} \fa_{ib}}
    \label{SM-Delta-summary2}
    \\
    [\B_i^j]_{ba} 
    &=(1-\delta_{ij})
    \left[\Pi_jf(h_*)\Pi_i\frac{1}{\sqrt{^t\Delta_i(\mathbf{1}-^t\Delta_i)}}\right]_{ba}
    \label{SM-B-summary2}
    \\
    [\D_i]_{b\alpha}&=\sum_{j=1}^{\mathcal{N}}
    \left[t_{ij}\R^\dagger_j\B^j_i\right]_{\alpha b}
    \label{SM-D-summary2}
    \\
    [\Lambda^c_i]_{ab}&=-[\Pi_i \B_i^\dagger h_* \B_i\Pi_i]_{ab}
    \label{SM-Lc-summary2}
    \\
    \hat{H}^i_{emb}\ket{\Phi_i}&=E^c_i\ket{\Phi_i}
    \label{SM-Hemb-summary2}
    \\
    \Av{\Phi_i}{\bbc_{ia}\bba_{ib}}&=[\mathbf{1}-\Delta_i]_{ab}
    \label{SM-SC-bath-summary2}
    \\
    \Av{\Phi_i}{\cc_{i\alpha}\bba_{ib}}&=\sum_{a=1}^{B\nu_i} [\R_{i}]_{a\alpha}
    [\Delta_i(\mathbf{1}-\Delta_i)]^{\frac{1}{2}}_{ab}
    \label{SM-SC-hybr-summary2}
    \,,
\end{align}
where:
\begin{align}
    \B_i &= \sum_{j=1}^{\mathcal{N}}{\B}_i^j
    \,.
\end{align}

We note that Eq.~\eqref{SM-Hqp-summary}, Eq.~\eqref{SM-Delta-summary}, Eq.~\eqref{SM-Hemb-summary} and Eq.~\eqref{detF1} are 
identical to Eq.~\eqref{SM-Hqp-summary2}, Eq.~\eqref{SM-Delta-summary2}, Eq.~\eqref{SM-Hemb-summary2} and Eq.~\eqref{SM-SC-hybr-summary2}, respectively.
The equivalence between Eq.~\eqref{detD} and Eq.~\eqref{SM-D-summary2} can be readily verified by inspection by substituting Eq.~\eqref{SM-B-summary2} in Eq.~\eqref{SM-D-summary2};
while the equivalence between Eq.~\eqref{detF2} and Eq.~\eqref{SM-SC-bath-summary2} is a trivial consequence of the canonical Fermionic anticommutation rules of the bath modes:
$\{\bba_{ia},\bbc_{ib}\}=\delta_{ab}$.

\subsection*{Proof of equivalence between Eq.~\eqref{detLc} and Eq.~\eqref{SM-Lc-summary2}}

Proving the equivalence between Eq.~\eqref{detLc} and Eq.~\eqref{SM-Lc-summary2} is not trivial, as Eq.~\eqref{detLc} involves directional derivatives of a function of the matrix $\Delta_i$ with respect to the matrices $^t\left[h_i\right]_s$, which do not generally commute
with $\Delta_i$.
To circumvent this technical problem we use the following indirect strategy.

We know that both the gDMET equations [Eqs.~\eqref{SM-Hqp-summary2}-\eqref{SM-SC-hybr-summary2}] for $B=1$ and 
the gGA equations [Eqs.~\eqref{SM-Hqp-summary}-\eqref{detF2}] for $B=1$ (i.e., the standard GA)
are sufficient for calculating \emph{exactly} the solution of any one-body Hamiltonian
$\hat{H}_0$, and in such case all of the respective self-consistency conditions are exactly satisfied by setting $\hat{H}_*=\hat{H}_0$.
Our proof is based on the idea of writing explicitly this condition for a one-body 
Hamiltonian represented as follows:
\begin{align}
    \hat{H}_0=\sum_{i,j=1}^{\mathcal{N}}\sum_{a,b=1}^{B{\nu}_i}\left[\R_i^\dagga t_{ij}\R^{\dagger}_j\right]_{ab}
    \fc_{ia}\fa_{jb}
    + \sum_{i=1}^{\mathcal{N}}\sum_{a,b=1}^{B{\nu}_i} \left[ \Lambda_i \right]_{ab}\fc_{ia} \fa_{ib}
    \label{SM-Hqp-proof}
\end{align}
where $\R,\Lambda$ are arbitrary \emph{fixed} parameters.

From the observation above it follows that the following equations are satisfied exactly:
\begin{align}
    \hat{H}_0\ket{\Psi_0}&=E_0\ket{\Psi_0}
    \label{SMproof-1}
    \\
    [{\Delta}_i]_{ab}&=\Av{\Psi_0}{\fc_{ia} \fa_{ib}}
    \label{Delta-proof}
    \\
    [\B_i^j]_{ba} 
    &=(1-\delta_{ij})
    \left[\Pi_jf(h_*)\Pi_i\frac{1}{\sqrt{^t\Delta_i(\mathbf{1}-^t\Delta_i)}}\right]_{ba}
    \\
    [{\D}^0_i]_{b a}&=\sum_{j=1}^{\mathcal{N}}
    \left[\R_it_{ij}\R^\dagger_j{\B}^j_i\right]_{a b}
    =\sum_{\alpha=1}^{\nu_i}[\D_i]_{b\alpha}[\R_i]_{a\alpha}
    \label{D-proof}
    \\
    [{\Lambda}^c_i]_{ab}&=-[\Pi_i {\B}_i^\dagger h_* {\B}\Pi_i]_{ab}
    \label{Lc-proof}
    \\
    \hat{H}^i_{emb}\ket{\Phi^0_i}&=E^c_i\ket{\Phi^0_i}
    \\
    \Av{{\Phi}^0_i}{\bbc_{ia}\bba_{ib}}&=[\mathbf{1}-{\Delta}_i]_{ab}
    \label{SC-bath-proof}
    \\
    \Av{{\Phi}^0_i}{\fc_{ia}\bba_{ib}}&=
    [{\Delta}_i(\mathbf{1}-{\Delta}_i)]^{\frac{1}{2}}_{ab}
    \label{SC-hybr-proof}
    \\
    \Av{{\Phi}^0_i}{\fc_{ia}\fa_{ib}}&=[{\Delta}_i]_{ab}
    \label{SC-imp-proof}
    \,,
\end{align}
where $h_*$ is defined as in Eqs.~\eqref{SM-h*},\eqref{SM-Proj}, and:
\begin{align}
    \hat{H}^i_{emb}&= \sum_{a,b=1}^{{\nu}_i}\left[\Lambda_i\right]_{ab}\fc_{ia}\fa_{ib}
    + \sum_{a=1}^{{\nu}_i}\sum_{\alpha=1}^{\nu_i}\left[\D^0_i\right]_{a\alpha}\cc_{i\alpha}\bbc_{ia}
   +\sum_{a,b=1}^{{\nu}_i}\left[\Lambda^c_i\right]_{ab}\bba_{ib}\bbc_{ia}
   \,.
   \label{SM-hemb-proof}
\end{align}
Note that Eq.~\eqref{Lc-proof} coincides exactly with Eq.~\eqref{SM-Lc-summary2}
(while Eq.~\eqref{D-proof} for $D^0_i$ is different from Eq.~\eqref{SM-D-summary2} for $D_i$).

Our proof arises from the observation that also the standard GA framework leads to Eqs.~\eqref{SMproof-1}-\eqref{SM-hemb-proof}, except for the seemingly different expression for the EH bath parameters ${\Lambda}^c_i$:
\begin{align}
[l^c_i]_{s}
   &=-[l_i]_{s}-\sum_{c,b=1}^{B{\nu}_i}\sum_{\alpha=1}^{\nu_i}\frac{\partial}{\partial \left[d^0_i\right]_s} \left(\left[\Delta_i\left(\mathbf{1}-\Delta_i\right)\right]^{\tfrac{1}{2}}_{cb}\left[\D^0_i\right]_{bc} + \mathrm{c.c.}\right)
   \nonumber\\
    &=-[l_i]_{s}-\sum_{c,b=1}^{B{\nu}_i}\sum_{\alpha=1}^{\nu_i}\frac{\partial}{\partial \left[d^0_i\right]_s} \left(\left[\Delta_i\left(\mathbf{1}-\Delta_i\right)\right]^{\tfrac{1}{2}}_{cb}\left[\D_i\right]_{b\alpha}\left[\R_i\right]_{c\alpha} + \mathrm{c.c.}\right) 
    \label{steplc}
    \,,
\end{align}
where the first line of Eq.~\eqref{steplc} stems from Eq.~\eqref{detLc}, considering that $\hat{H}_*=\hat{H}_0$ for the one-body system here considered for our proof; while the second step stems from the fact that
$[{\D}^0_i]_{b a}=\sum_{\alpha=1}^{\nu_i}[\D_i]_{b\alpha}[\R_i]_{a\alpha}$,
as noted on the right hand side of Eq.~\eqref{D-proof}.

This implies that the gDMET and gGA expressions for $\hat{H}^i_{emb}$ must have the
same $\Lambda_i$ and $\D^0_i$, while they may (in principle) have different $\Lambda^c_i$.
On the other hand, whether we use the gDMET construction or the gGA construction, the entries of the ground-state single-particle density matrix
of the embedding Hamiltonian are given by [Eqs.~\eqref{SC-bath-proof}-\eqref{SC-imp-proof}], i.e., they must be the same in both approaches.
In conclusion, provided that there not exist multiple possible bath parameters providing the same ground-state single-particle density matrix of the EH (at fixed $\Lambda_i$ and ${\D}_i^0$),
the 2 seemingly different expressions for $\Lambda^c_i$ given by Eqs.~\eqref{steplc} and
\eqref{Lc-proof} must be equivalent.
On the other hand, Eq.~\eqref{steplc} is preferable for practical implementations, both because of its simplicity and because of its lower computational cost.

\section{The $\text{g}$DMET equations for translationally invariant systems}

For completeness, in this section we write explicitly the gDMET equations [Eqs.~\eqref{SM-Hqp-summary2}-\eqref{SM-SC-hybr-summary2}] for systems with translational symmetry, represented as:
\begin{align}
    \hat{H}&=\sum_{\bR}\sum_{i}\hat{H}_{loc}^i[\cc_{\bR i\alpha},\ca_{\bR i\alpha}]+
    \sum_{\bk}\sum_{ij}\sum_{\alpha=1}^{\nu_i}\sum_{\beta=1}^{\nu_j}[t_{\bk,ij}]_{\alpha\beta}\,\cc_{\bk i\alpha}\ca_{\bk j\beta}  
    \label{H-k}
    \\
    &=\sum_{\bR}\sum_{i}\hat{H}_{loc}^i[\cc_{\bR i\alpha},\ca_{\bR i\alpha}]+
    \sum_{\bR,\bRp}\sum_{ij}\sum_{\alpha=1}^{\nu_i}\sum_{\beta=1}^{\nu_j}[t_{\bR i,\bRp j}]_{\alpha\beta}\,\cc_{\bR i\alpha}\ca_{\bRp j\beta}  
    \,,
    \label{H-real}
\end{align}
where $\bk$ is the momentum conjugate to the unit-cell label $\bR$, the fragments within each unit cells
are labeled by $i,j$, the corresponding spin-orbitals are labeled by $\alpha,\beta$, and:
\begin{align}
t_{\bk,ij}=\Pi_it_{\bk}\Pi_j    
\end{align} 
is the hopping matrix expressed in momentum representation, which is related to the
real-space representation as follows:
\begin{align}
    t_{\bR i,\bRp j} = \frac{1}{\mathcal{N}}\sum_{\bk} e^{i\bk(\bR-\bRp)} t_{\bk,ij}
    \,,
\end{align}
where $\mathcal{N}$ is the number of $\bk$ points.
As in previous work~\cite{SM-Our-PRX}, with no loss of generality, we assume that the second term in Eqs.~\eqref{H-k} and \eqref{H-real}
is nonlocal, i.e., that:
\begin{align}
    \sum_{\bk} t_{\bk,ii}=0\qquad\forall\,i\,,
\end{align}
and all one-body and two-body local terms of the Hamiltonian are included in $\hat{H}_{loc}^i$.

Using the definition above, Eqs.~\eqref{SM-Hqp-summary2}-\eqref{SM-SC-hybr-summary2} can be rewritten as follows:
\begin{align}
    \hat{H}_*[\R,\Lambda]\ket{\Psi_0}&=E_0\ket{\Psi_0}
    \label{SM-Hqp-summary3}
    \\
    [{\Delta}_i]_{ab}&=\frac{1}{\mathcal{N}}\sum_{\bk}\Av{\Psi_0}{\fc_{\bk ia} \fa_{\bk ib}}
    \label{SM-Delta-summary3}
    \\
    [{\B}_i^j(\bk)]_{b a} &= 
    \left[
    \Pi_{j}\left(
    f(t^*_\bk)-\delta_{ij}\,^t\Delta_i)\Pi_{i}
    \right)
    \frac{1}{\sqrt{^t\Delta_i(\mathbf{1}-^t\Delta_i)}}\right]_{ba}
    \label{SM-B-summary3}
    \\
    [\D_i]_{b\alpha}&=\frac{1}{\mathcal{N}}\sum_{\bk,j}
    \left[\Pi_i
    t_{\bk}\Pi_j\R_j^\dagger\Pi_j\B^j_i(\bk)\right]_{\alpha b}
    \label{SM-D-summary3}
    \\
    [\Lambda^c_i]_{ab}&=-\frac{1}{\mathcal{N}}\sum_{\bk}\sum_{jj'}
    \left[
    \Pi_i [\B_i^j(\bk)]^\dagger
    \Pi_j t_*(\bk) \Pi_{j'}\B^{j'}_i(\bk)\Pi_i
    \right]_{ab}
    \label{SM-Lc-summary3}
    \\
    \hat{H}^i_{emb}\ket{\Phi_i}&=E^c_i\ket{\Phi_i}
    \label{SM-Hemb-summary3}
    \\
    \Av{\Phi_i}{\bbc_{ia}\bba_{ib}}&=[\mathbf{1}-\Delta_i]_{ab}
    \label{SM-SC-bath-summary3}
    \\
    \Av{\Phi_i}{\cc_{i\alpha}\bba_{ib}}&=\sum_{a=1}^{B\nu_i} [\R_{i}]_{a\alpha}
    [\Delta_i(\mathbf{1}-\Delta_i)]^{\frac{1}{2}}_{ab}
    \label{SM-SC-hybr-summary3}
    \,,
\end{align}
where $\hat{H}^i_{emb}$ is given by Eq.~\eqref{SM-hemb}, while:
\begin{align}
    \hat{H}_*&=\sum_\bk \sum_{i,j}\sum_{a=1}^{B\nu_i}\sum_{b=1}^{B\nu_j}
    \left[\R_i^\dagga t_{\bk,ij}\R^{\dagger}_j\right]_{ab}
    \fc_{\bk ia}\fa_{\bk jb}
    + \sum_{\bk}\sum_{i}\sum_{a,b=1}^{B{\nu}_i} \left[ \Lambda_i \right]_{ab}\fc_{\bk ia} \fa_{\bk ib}
    \\
    t^*_{\bk,ij}&=\Pi_i t^*_{\bk} \Pi_j=\R_i\Pi_it_{\bk}\Pi_j\R_j^\dagger
    \,.
\end{align}


%

\end{document}